\documentclass[submission, Phys]{SciPost}

\usepackage{amsmath,amssymb, bbm, dsfont}
\usepackage{graphicx}
\usepackage[nopatch=footnote]{microtype}
\usepackage{xcolor}
\usepackage{grffile}
\usepackage[normalem]{ulem}
\usepackage{cancel}
\usepackage{bm}
\usepackage{setspace}
\usepackage{enumitem}
\usepackage{lineno}

\makeatletter
\@ifpackageloaded{hyperref}{%
  \hypersetup{colorlinks=true,linkcolor=blue,urlcolor=blue,citecolor=red}%
}{}
\makeatother

\newcommand{\bra}[1]{\langle #1|}
\newcommand{\ket}[1]{|#1\rangle}
\newcommand{\average}[1]{\langle #1\rangle}

\newcommand{\dbra}[1]{\langle\!\langle #1}
\newcommand{\dket}[1]{|#1\rangle\!\rangle}

\def\id{\mathds{1}}

\makeatletter
\providecommand{\affiliation}[1]{}
\providecommand{\email}[1]{}
\providecommand{\maketitle}{}

\renewcommand{\title}[1]{}
\renewcommand{\author}[1]{}
\renewcommand{\affiliation}[1]{}
\renewcommand{\email}[1]{}
\renewcommand{\date}[1]{}
\renewcommand{\maketitle}{}

\@ifundefined{abstract}{%
  \newenvironment{abstract}
    {\section*{Abstract}\begingroup\bfseries}
    {\endgroup
     \vspace{10pt}
     \noindent\rule{\textwidth}{1pt}
     \tableofcontents\thispagestyle{fancy}
     \noindent\rule{\textwidth}{1pt}
     \vspace{10pt}}
}{%
  \renewenvironment{abstract}
    {\section*{Abstract}\begingroup\bfseries}
    {\endgroup
     \vspace{10pt}
     \noindent\rule{\textwidth}{1pt}
     \tableofcontents\thispagestyle{fancy}
     \noindent\rule{\textwidth}{1pt}
     \vspace{10pt}}
}

\@ifundefined{acknowledgments}{%
  \newenvironment{acknowledgments}{\section*{Acknowledgements}}{}
}{%
  \renewenvironment{acknowledgments}{\section*{Acknowledgements}}{}
}
\makeatother

\begin{document}

\begin{center}{\Large \textbf{
Dynamical correlations in a dissipative XXZ spin chain
}}\end{center}

\begin{center}
C\u at\u alin ~Pa\c scu Moca\textsuperscript{1,2,3*},
Doru Sticlet\textsuperscript{4},
Ovidiu I. P\^ a\c tu\textsuperscript{5},
Balazs D\'ora\textsuperscript{2,3}
\end{center}
\affiliation{Institute for Space Sciences, Bucharest-M\u agurele, R 077125, Romania}
\begin{center}
{\bf 1} Department of Physics, University of Oradea, 410087, Oradea, Romania
\\
{\bf 2} Department of Theoretical Physics, Institute of Physics, Budapest University of Technology and Economics, M\H{u}egyetem rkp.~3, H-1111 Budapest, Hungary
\\
{\bf 3} MTA-BME Lend\"ulet ``Momentum'' Open Quantum Systems Research Group, Institute of Physics, Budapest University of Technology and Economics, M\H{u}egyetem rkp.~3, H-1111 Budapest, Hungary
\\
{\bf 4} National Institute for R\&D of Isotopic and Molecular Technologies, 67-103 Donat, 400293 Cluj-Napoca, Romania
\\
{\bf 5} Institute for Space Sciences, Bucharest-M\u agurele, R 077125, Romania
\\
* Corresponding author: mocap@uoradea.ro
\end{center}

\begin{center}
\today
\end{center}


\begin{abstract}
We study dynamical spin correlations in a dissipative XXZ spin chain subject to uniform local spin-loss and pumping. Starting from a 
mixed steady state that is featureless albeit possessing finite magnetization, rich dynamics emerges in time-dependent two-point correlators evaluated on top of it. 
For unitary evolution in which the reservoir is absent, the longitudinal correlators reproduce the established hierarchy 
of spin-transport universality classes — ballistic, Kardar--Parisi--Zhang (KPZ) superdiffusive, and diffusive — across the phase diagram. 
However, for finite magnetization, additional ballistic light cone propagation gets
superimposed on the previous universality classes, arising from magnon propagation.
The transverse correlator displays very fast, exponential decay of correlations without wavefront propagation in the easy-plane case. 
At the isotropic point, it follows KPZ scaling due to $SU(2)$ symmetry,
while in the easy-axis regime, it is characterized by ballistic spreading of correlations.
Under full Lindbladian dynamics, the universality classes are preserved at early times, while the correlations acquire an overall exponential 
damping in the long-time limit. In terms of methods, we have used vectorized TEBD for numerical simulations and exact analytical results 
obtained via a Pfaffian representation and the third-quantization framework for the noninteracting XX case.
\end{abstract}

\section{Introduction}\label{sec:introduction}

Understanding the dynamics of strongly correlated quantum systems far from equilibrium is one of the central challenges 
of modern condensed matter and quantum information physics. 
While the thermalization of isolated quantum systems is relatively
well established through the eigenstate thermalization hypothesis~\cite{rigol2008thermalization, 
polkovnikov2011colloquium}, real experiments invariably involve coupling to an environment, making Markovian open 
quantum systems a natural and increasingly important paradigm. Driven by advances in ultracold atomic gases~\cite{bloch2008many, gross2017quantum}, trapped ions~\cite{barreiro2011open}, and superconducting circuits~\cite
{verstraete2009quantum, sieberer2016keldysh}, these platforms now offer exquisite control over both coherent and 
incoherent processes, motivating  questions about how dissipation modifies the fundamental dynamical properties 
of strongly interacting quantum systems.

The one-dimensional spin-$1/2$ XXZ chain is an ideal testing ground for such questions. Its equilibrium physics is 
exactly solvable by the Bethe ansatz~\cite{yang1966one, takahashi1999thermodynamics, giamarchi2003quantum} and its 
spin-transport universality classes are quantitatively understood within the framework of generalized hydrodynamics 
(GHD)~\cite{castroalvaredo2016emergent, bertini2016transport, caux2016quench, doyon2020lecture}: transport is ballistic 
in the easy-plane phase ($|\Delta|<1$), crosses over to Kardar--Parisi--Zhang (KPZ) superdiffusion at the isotropic 
Heisenberg point ($\Delta=1$)~\cite{ljubotina2019kardar, ilievski2018superdiffusion, nardis2020superdiffusion, bulchandani2021kardar},
 and becomes diffusive in the easy-axis phase ($|\Delta|>1$)~\cite{zotos1997transport, bertini2021finite}.

When coupled to a Markovian bath via incoherent spin-flip processes at every site, 
the chain is described by the Lindblad master 
equation~\cite{lindblad1976generators, gorini1976completely, breuer2002theory} and is driven to a unique nonequilibrium 
steady state (SS) that is exactly solvable for any $\Delta$~\cite{prosen2011open, znidaric2011spin}. 
Remarkably, this SS is a featureless product state, without any correlations, governed solely by the ratio of pumping to loss rates — the spin 
fugacity $\zeta$ — and completely independent of $J$ and $\Delta$~\cite{prosen2012diffusive}. 
Despite vanishing static 
correlations, nontrivial dynamics appears in time-dependent two-point correlators evaluated on top of the SS, which 
are accessible via the quantum regression theorem~\cite{lax1963formal, carmichael1993open} and probe the interplay of 
coherent spin transport and incoherent decoherence~\cite{alba2021spreading, turkeshi2021diffusion}. 

In the present work we address this problem and compute longitudinal and transverse spin correlators across the full phase diagram and for different values of the fugacity $\zeta$, for both unitary and Lindbladian evolution starting from the SS.
The purely unitary dynamics can be interpreted as a quantum quench protocol, where the system is first driven to the SS of the 
full open problem and then the dissipation is suddenly switched off, allowing us to probe the universal transport physics of the 
XXZ chain starting from a well-controlled, initial mixed state. 
Under unitary evolution, we find that the longitudinal channel reproduces the GHD hierarchy of universality classes, namely ballistic, KPZ, and diffusive scalings. 
For finite initial magnetization, an additional ballistic light cone propagation appears due to the free propagation of finite density of magnons~\cite{weiner2020}.

The transverse channel  behaves qualitatively differently: it decays exponentially without a light cone 
for $|\Delta|<1$, shows KPZ algebraic decay at $\Delta=1$, and supports a sharp ballistic front for $|\Delta|>1$. In addition,
the transverse correlators are largely unaffected by the finite initial magnetization, which is in the longitudinal direction.
We find that KPZ universality at $\Delta=1$ persists for all simulated fugacities, consistent with its protection by the $SU(2)$ symmetry~\cite{krajnik2020kardar, moca2023kardar}.

Under Lindbladian dynamics, the central result is that the light-cone structure of all three transport universality classes is 
preserved by bulk dissipation: the shape of the light cone remains
 unchanged up to an effective crossover time scale $t^{*}_{\rm eff} \sim 1/\Gamma_{\rm eff}$.
For $t\ll t^{*}_{\rm eff}$ coherent exchange dominates and the causal structure is fully intact; for $t\gtrsim t^{*}_{\rm eff}$ dissipation imposes an exponential decay in the correlations that makes the light-cone front irrelevant.
In the noninteracting limit $\Delta=0$, the effective rate reduces to $\Gamma_{\rm eff} = \Gamma\equiv \gamma_l + \gamma_p$ and acquires distinct $\Delta$- and $\zeta$-dependent corrections in the presence of interactions. Numerically, we 
employ vectorized TEBD~\cite{zwolak2004mixed, paeckel2019time, moca2022simulating}, while exact results for $\Delta=0$ 
are derived using a Pfaffian representation and the third-quantization ($3^{\mathrm{rd}}$QT) framework~\cite{prosen2008third, weimer2021simulation}.

The rest of the paper is organized as follows.
Sec.~\ref{sec:model} introduces the model and its exact steady state.
Sec.~\ref{sec:correlations} defines the two-point correlators and the numerical methods employed.
Sec.~\ref{sec:longitudinal_unitary} and Sec.~\ref{sec:transverse_unitary} present results for unitary evolution in the longitudinal and transverse channels, respectively.
Sec.~\ref{sec:Lindblad_longitudinal} and Sec.~\ref{sec:transversal_Lindblad} extend the analysis to full Lindbladian dynamics.
Sec.~\ref{sec:conclusions} summarizes the conclusions.
Appendix~\ref{sec:Majorana} provides the Majorana covariance matrices, Appendix~\ref{sec:3rdQT} presents the third-quantization formalism, Appendix~\ref{sec:Pfaffian} details the Pfaffian formulation of the transverse correlator, and Appendix~\ref{sec:TEBD} details the vectorized TEBD simulations for correlation functions.

\section{Dissipative XXZ model}\label{sec:model}

We consider a chain of $L$ spin-$1/2$ sites, subject to spatially uniform incoherent spin-flip processes. The coherent dynamics is governed by the XXZ Hamiltonian
\begin{equation}\label{eq:XXZ}
H = {J\over 4} \sum_{l=-L/2}^{L/2-1}
\left(
\sigma_l^x \sigma_{l+1}^x
+ \sigma_l^y \sigma_{l+1}^y
+ \Delta \, \sigma_l^z \sigma_{l+1}^z
\right),
\end{equation}
where $\sigma_l^{\alpha}$ ($\alpha = x,y,z$) are Pauli matrices at site $l$, $J>0$ is the exchange energy, and $\Delta$ is the dimensionless anisotropy parameter.

The coupling to a Markovian environment is described by the Lindblad master equation~\cite{lindblad1976generators, gorini1976completely}
\begin{eqnarray}
\frac{\mathrm{d}\rho(t)} {\mathrm{d}t} &=&  -i \mathcal{L}[\rho(t)]\nonumber\\
                                        &=& -i [H,\rho(t)] + \mathcal{D}[\rho(t)],
\label{eq:lindblad}
\end{eqnarray}
which provides the most general Markovian, completely positive, and trace-preserving map for the density matrix $\rho(t)$~\cite{breuer2002theory}. 
The coherent part generates unitary evolution, while $\mathcal{D}[\rho]$ encodes the action of the bath.

We focus on spin loss and pumping applied uniformly at every site, a dissipation channel realizable in quantum simulation platforms based on ultracold atoms in optical lattices~\cite{diehl2008quantum, barreiro2011open} and in circuit QED systems~\cite{verstraete2009quantum, sieberer2016keldysh}. The jump operators are
\begin{equation}
L_l^{-} = \sqrt{\gamma_l} \, \sigma_l^{-},
\qquad
L_l^{+} = \sqrt{\gamma_p} \, \sigma_l^{+},
\end{equation}
with $\sigma_l^{\pm} = (\sigma_l^x \pm i \sigma_l^y)/2$. Here $\gamma_l \geq 0$ is the single-site spin-loss rate and $\gamma_p \geq 0$ is the incoherent pumping rate. The total single-site decoherence rate $\Gamma = \gamma_l + \gamma_p$ sets the  dissipative time scale $t^{*}=\Gamma^{-1}$ in the noninteracting limit, $\Delta=0$. The Lindblad dissipator takes the form
\begin{gather}
\mathcal{D}[\rho]
=
\sum_{l=-L/2}^{L/2}
\Big[
\gamma_l
\big(
\sigma_l^{-} \rho \sigma_l^{+}
- \tfrac{1}{2} \{ \sigma_l^{+}\sigma_l^{-}, \rho \}
\big)
+\gamma_p
\big(
\sigma_l^{+} \rho \sigma_l^{-}
- \tfrac{1}{2} \{ \sigma_l^{-}\sigma_l^{+}, \rho \}
\big)
\Big].
\label{eq:Dissipation}
\end{gather}
The two terms describe, respectively, transitions from $|\!\uparrow\rangle$ to $|\!\downarrow\rangle$ (spin loss, rate $\gamma_l$) and from $|\!\downarrow\rangle$ to $|\!\uparrow\rangle$ (pumping, rate $\gamma_p$). The competition between these processes and the coherent XXZ exchange drives the system to a nonequilibrium steady state~\cite{prosen2008third, prosen2011open, znidaric2011spin},
The Lindblad dynamics possesses a unique SS, defined as the density matrix $\rho_{\mathrm{SS}}$ that is annihilated by the Lindblad superoperator,
\begin{equation}
\mathcal{L}[\rho_{\mathrm{SS}}] = 0.
\label{eq:SS}
\end{equation}
The key dimensionless parameter governing its structure is the pumping-to-loss ratio
\begin{equation}
\zeta = \frac{\gamma_p}{\gamma_l},
\end{equation}
which we refer to as the spin fugacity. Remarkably, the SS can be constructed exactly for any values of $\zeta$, $J$, and $\Delta$~\cite{prosen2011open}: it corresponds to a grand-canonical infinite-temperature state at fixed fugacity $\zeta$, completely insensitive to the exchange coupling and anisotropy. Explicitly,
\begin{equation}\label{eq:rhoSS}
\rho_{\mathrm{SS}}
=
\frac{1}{(1+\zeta)^L}
\sum_{M=0}^{L}
\zeta^M \, \mathds{1}_M,
\end{equation}
where $\mathds{1}_M$ is the identity restricted to the sector with total magnetization $M$.
The physical mechanism behind this exact result is transparent. The state $\rho_{\mathrm{SS}}$ is diagonal in the $\sigma^z$ eigenbasis and depends on a spin configuration only through the total magnetization $M$. Since the XXZ Hamiltonian conserves $S^z_{\mathrm{tot}} = \frac{1}{2}\sum_l \sigma_l^z$, any density matrix that is a function of $S^z_{\mathrm{tot}}$ alone commutes with $H$, so $[H, \rho_{\mathrm{SS}}]=0$. Independently, the local balance condition $\gamma_l \zeta = \gamma_p$ ensures that each single-site dissipator annihilates the local  $\rho_l^{\mathrm{SS}} = (\zeta|\!\uparrow\rangle\langle\uparrow\!| + |\!\downarrow\rangle\langle\downarrow\!|)/(1+\zeta)$, so $\mathcal{D}[\rho_{\mathrm{SS}}]=0$, which  guarantee that $\mathcal{L}[\rho_{\mathrm{SS}}]=0$.
The SS takes an equivalent and more transparent form as a completely factorized product state,
\begin{equation}
\rho_{\mathrm{SS}}
=
\frac{1}{(1+\zeta)^L}
\bigotimes_{l=-L/2}^{L/2}
\begin{pmatrix}
\zeta & 0 \\
0 & 1
\end{pmatrix},
\label{eq:rho_NESS}
\end{equation}
with the local basis $\{ \ket{\uparrow}, \ket{\downarrow} \}$, revealing explicitly that $\rho_{\mathrm{SS}}$ contains no intersite 
entanglement or correlations. 
All connected equal-time spin-spin correlation functions therefore vanish identically in the SS. 
The expression~\eqref{eq:rho_NESS}  
is particularly useful for numerical simulations, as it allows us to prepare the SS as a simple product state as an MPO  and then probe 
the dynamical generation of correlations on top of it.
The uniform local magnetization is
\begin{equation}
\langle \sigma_l^z \rangle_{\mathrm{SS}} = \frac{\zeta-1}{\zeta+1},
\label{eq:Sz_NESS}
\end{equation}
ranging from $-1$ ($\zeta\to 0$, fully polarized down) to $+1$ ($\zeta\to\infty$, fully polarized up), while transverse coherences vanish identically, $\langle \sigma_l^{\pm} \rangle_{\mathrm{SS}} = 0$.
\begin{figure}[tbh!]
  \begin{center}
   \includegraphics[width=0.55\textwidth]{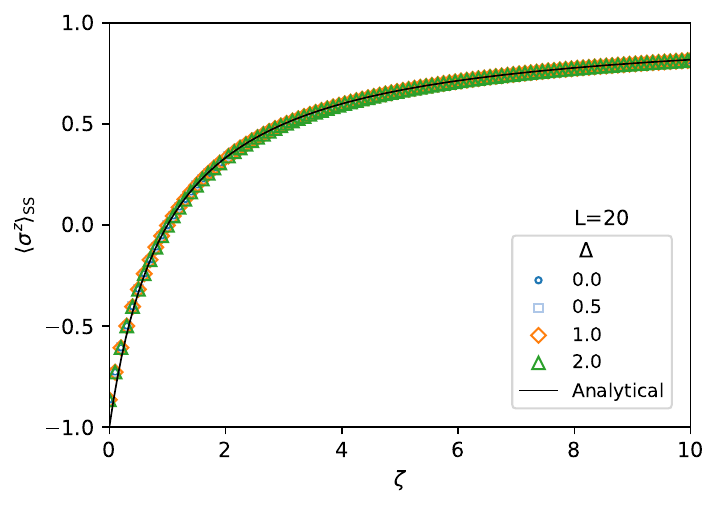}
   \caption{Average local magnetization $\langle\sigma_l^z\rangle_{\mathrm{SS}}$ versus fugacity $\zeta$ for several anisotropy parameters $\Delta$, obtained with the vectorized TEBD method. All symbols collapse onto the analytical prediction~Eq.~\eqref{eq:Sz_NESS} (solid line), confirming that the SS structure is governed entirely by the gain--loss balance $\zeta = \gamma_p/\gamma_l$ and is independent of $J$ and $\Delta$. System size is $L=20$ and bond dimension $\chi=128$. The results have been checked for convergence in both parameters.}
   \label{fig:NESS}
  \end{center}
\end{figure}

Figure~\ref{fig:NESS} benchmarks the SS against vectorized TEBD simulations for different values of $\Delta$. The numerical data collapse 
exactly onto Eq.~\eqref{eq:Sz_NESS} for all anisotropies, confirming that coherent exchange interactions play no role in determining the 
steady-state magnetization and that the SS structure is set entirely by the dissipative rates through $\zeta$ and  independent of $\Delta$.

At the special point $\zeta=1$ ($\gamma_l = \gamma_p$), the SS reduces to the infinite-temperature state, $\rho_{\mathrm{SS}} \propto 
\mathds{1}$, with vanishing average magnetization $\langle\sigma^z_l\rangle_{\mathrm{SS}} = 0$. 
Although static correlations are completely suppressed in the SS for any $\zeta$, nontrivial physics emerges in the time-dependent correlation functions of operators evaluated on top of this state. 
These dynamical correlators probe how spin excitations propagate and relax in the presence of coherent 
XXZ exchange and local dissipation, and form the central object of study in this work.

\section{Correlation functions}\label{sec:correlations}

Motivated by the factorized, correlation-free structure of the SS established in Sec.~\ref{sec:model}, we turn to the central objects of this work: time-dependent two-point spin correlation functions evaluated on top of $\rho_{\mathrm{SS}}$. 
Such dynamical correlators are the natural probes of spin excitations, transport, and the interplay between coherent and dissipative dynamics in driven open systems~\cite{sieberer2016keldysh, breuer2002theory, damanet2019atom}. 
The primary object of study is the two-time, two-point correlator under the \emph{full} Lindbladian evolution,
\begin{equation}
\average{\sigma_l^{\alpha}(t)\,\sigma_{l'}^{\beta}}^{(\mathcal{L})}
=
\mathrm{Tr}\!\left\{
\sigma_l^{\alpha}\,
e^{-i\mathcal{L} t}
\big[
\sigma_{l'}^{\beta}\,\rho_{\mathrm{SS}}
\big]
\right\},
\label{eq:C_L_spin}
\end{equation}
with $\alpha,\beta\in\{x,y,z\}$. This quantity follows directly from the quantum regression theorem~\cite{lax1963formal, carmichael1993open}, which relates two-time correlators of Markovian open systems to the propagation of an 
auxiliary density matrix under the same Lindbladian. Equation~\eqref{eq:C_L_spin} therefore probes how a local spin disturbance, 
imprinted on the SS by the action of $\sigma_{l'}^{\beta}$ at $t=0$, propagates and relaxes under the combined action of coherent XXZ exchange and local incoherent spin flips.

For comparison, and to isolate the role of dissipation, we define the \emph{unitary} correlator
\begin{eqnarray}
\average{\sigma_l^{\alpha}(t)\,\sigma_{l'}^{\beta}}^{(\mathcal{H})}
&=&
\mathrm{Tr}\!\left\{
\sigma_l^{\alpha}\,
e^{-i\mathcal{L}_0 t}
\big[
\sigma_{l'}^{\beta}\,\rho_{\mathrm{SS}}
\big]
\right\}  = 
\mathrm{Tr}\!\left\{
e^{iHt}\,\sigma_l^{\alpha}\,e^{-iHt}\,
\sigma_{l'}^{\beta}\,\rho_{\mathrm{SS}}
\right\},
\label{eq:C_H_spin}
\end{eqnarray}
where $\mathcal{L}_0[\cdot]=[H,\cdot]$ generates purely unitary dynamics. This correlator can be 
interpreted as a quantum quench protocol~\cite{polkovnikov2011colloquium, essler2016quench}: the system is 
first driven to the SS, $\rho_{\mathrm{SS}}$ of the full open problem, and the dissipation is then 
suddenly switched off, after which the chain evolves under the closed XXZ Hamiltonian. This construction is 
relevant because it decouples the effect of the initial state (which encodes the fugacity $\zeta$ through 
the local spin polarization) from the coherent many-body dynamics, allowing us to access the universal 
transport physics of the XXZ chain starting from a well-controlled, analytically known initial mixed state.
The two classes of correlators considered in this work are therefore defined by the same initial state $\rho_{\mathrm{SS}}$ but differ in the presence or absence of dissipation during the time evolution.
For each class of correlators, we focus on the longitudinal ($\alpha=\beta=z$) and transverse ($\alpha=\beta=x$) channels, which probe distinct physical properties of the system and are expected to exhibit qualitatively different behavior:

\textit{(i) Longitudinal correlations.} --- The longitudinal spin-spin correlator $\average{\sigma_l^{z}(t)\,\sigma_{l'}^{z}}^{(\mathcal{L}/\mathcal{H})}$ probes fluctuations of the locally conserved magnetization 
density. Since $H$ conserves $S^z_{\mathrm{tot}}$ and the SS is diagonal in the $\sigma^z$ basis, 
equal-time connected longitudinal correlations vanish identically, $\average{\sigma_l^{z}\,\sigma_{l'}^{z}}_c^{\mathrm{SS}} = 0$ for $l\neq l'$ (subscript $c$ denotes the connected part). Nontrivial longitudinal correlations are therefore a purely dynamical 
signature, generated by the coherent spreading of magnetization fluctuations away from the steady state~\cite{bertini2021finite}. Their late-time behavior is controlled by spin transport and characterized by a 
universal power-law decay governed by the dynamical exponent $z$ of the relevant hydrodynamic mode~\cite{spohn2020fluctuating, ilievski2018superdiffusion, deNardis2019anomalous}.

\textit{(ii) Transverse correlations.} --- The transverse correlator $\average{\sigma_l^{x}(t)\,\sigma_{l'}^{x}}^{(\mathcal{L}/\mathcal{H})}$ probes spin coherence, i.e., the off-diagonal elements of the density 
matrix in the $\sigma^z$ basis. Although transverse coherences vanish locally in the SS, 
$\langle\sigma_l^\pm\rangle_{\mathrm{SS}}=0$, finite-time transverse correlations between different sites are 
dynamically generated by the coherent XXZ exchange and encode the spreading of quantum information along the 
chain~\cite{calabrese2005evolution}. 
At the isotropic point $\Delta=1$, $SU(2)$ symmetry ties the 
longitudinal and transverse correlators together. 
Transverse correlations therefore inherit the KPZ superdiffusive behavior~\cite{ljubotina2019kardar, moca2023kardar}. 

Both correlators are evaluated numerically within the matrix product state (MPS) framework~\cite{schollwock2011density, verstraete2008matrix, paeckel2019time}. 
For the Lindbladian correlator~\eqref{eq:C_L_spin} we use the vectorization of the density matrix~\cite{prosen2008third, moca2022simulating}: mapping $\rho(t)$ to a state $|\rho(t)\rangle\!\rangle$ in the doubled Hilbert space $\mathcal{H}\otimes\mathcal{H}$, the Lindblad superoperator $\mathcal{L}$ becomes a non-Hermitian effective Hamiltonian 
acting on this enlarged space~\cite{sticlet2026}. The quantum regression theorem then reduces the two-time correlator~\eqref{eq:C_L_spin} to propagating the auxiliary state $\sigma_{l'}^{\beta}|\rho_{\mathrm{SS}}\rangle\!\rangle$ 
with the vectorized Lindbladian, carried out via the time-evolving block decimation (TEBD) algorithm~\cite{vidal2007classical, paeckel2019time} [see Appendix~\ref{sec:TEBD}]. 
For the unitary correlator~\eqref{eq:C_H_spin}, the same MPS framework 
is applied without the dissipative part, which allows significantly larger accessible times and system sizes.

All simulations employ chains of length $L=129$ and bond dimensions $\chi=256$, with 
Trotter time step $\delta t = 0.01\,J^{-1}$. 
The results have been checked for convergence in both $\chi$ and $\delta t$.
In the noninteracting limit $\Delta=0$, the quadratic nature of both the Hamiltonian and the jump operators allows an exact treatment after a Jordan--Wigner 
mapping~\cite{prosen2008third}. This yields closed-form expressions for the longitudinal correlators under Lindbladian evolution and for the unitary XX correlators, 
providing benchmarks for the TEBD and third-quantization ($3^{\mathrm{rd}}$QT) calculations and clarifying which features are modified once interactions are 
turned on.

\section{Unitary evolution, Longitudinal correlations}\label{sec:longitudinal_unitary}

We start our analysis of the longitudinal spin-spin correlator under unitary (post-quench) evolution,
$\average{\sigma_l^{z}(t)\,\sigma_{l'}^{z}}^{(\mathcal{H})}$, with the system initialized in the $\rho_{\mathrm{SS}}$.
Since total magnetization $S^z_{\mathrm{tot}}$ is conserved, its late-time dynamics is controlled by spin
transport~\cite{zotos1997transport,bertini2021finite} and characterized by a dynamical exponent $z$.

\subsection{Noninteracting limit, $\Delta=0$}\label{sec:Delta_0}

In the noninteracting limit $\Delta=0$, the XXZ Hamiltonian reduces to the free-fermion XX model, which is solved exactly via a Jordan--Wigner transformation. The resulting fermionic Hamiltonian reads
\begin{equation}
H_{\mathrm{XX}} = \frac{J}{2}\sum_{l=-L/2}^{L/2-1}
\left( c^{\dagger}_l c_{l+1} + \mathrm{h.c.} \right),
\label{eq:H_spinless}
\end{equation}
and is diagonalized straightforwardly by a Fourier transform, yielding the dispersion relation $\epsilon(k)=J\cos k$.
The time evolution of the fermionic annihilation operators follows as
\begin{equation}
c_l(t) =
\sum_{l'=-\infty}^{\infty}
i^{\,l'-l}\,
J_{l'-l}(\Omega t)\,
c_{l'},
\end{equation}
where $\Omega = J$ and $J_n(x)$ denotes the Bessel function of the first kind. 
This representation makes explicit the ballistic spreading of single-particle excitations,
with a light-cone structure governed by the group velocity of the dispersion.
For a finite chain, this expression describes the bulk dynamics in the thermodynamic limit (or, equivalently, 
for times short enough that boundary reflections are negligible).
The initial state is given by Eq.~\eqref{eq:rho_NESS}. 
Correspondingly, the average occupation is
\begin{equation}
\bar{n} 
=
\average{c_l^{\dagger}c_l}
=
\frac{\zeta}{(\zeta+1)},
\label{eq:average_n}
\end{equation}
which is independent of position.
Using the Jordan--Wigner mapping, the longitudinal spin operator can be 
expressed as $\sigma_l^z = 2 c_l^{\dagger} c_l - 1$. 
Its time evolution therefore involves bilinear fermionic operators,
\begin{equation}
\sigma_l^{z}(t)
=
2\sum_{l',l''}
i^{\,l''-l'}
J_{l'-l}(\Omega t)
J_{l''-l}(\Omega t)\,
c_{l'}^{\dagger} c_{l''}
-1.
\end{equation}
Evaluating expectation values with respect to the initial steady state then yields an explicit expression for the longitudinal spin-spin correlation function under unitary evolution,
\begin{eqnarray}
\average{\sigma_0^{z}(t)\,\sigma_{0}^{z}}^{(\mathcal{H})}
&=&
(2\bar{n}-1)^2 + 4\bar{n}(1-\bar{n})\, J_{0}^{2}(\Omega t)
\nonumber\\
&=&
\langle \sigma^z \rangle_{\mathrm{SS}}^2
+\big(1-\langle \sigma^z \rangle_{\mathrm{SS}}^2\big)\,J_0^2(Jt).
\label{eq:S0_zz}
\end{eqnarray}
\begin{figure}[tbh!]
  \begin{center}
   \includegraphics[width=0.46\textwidth]{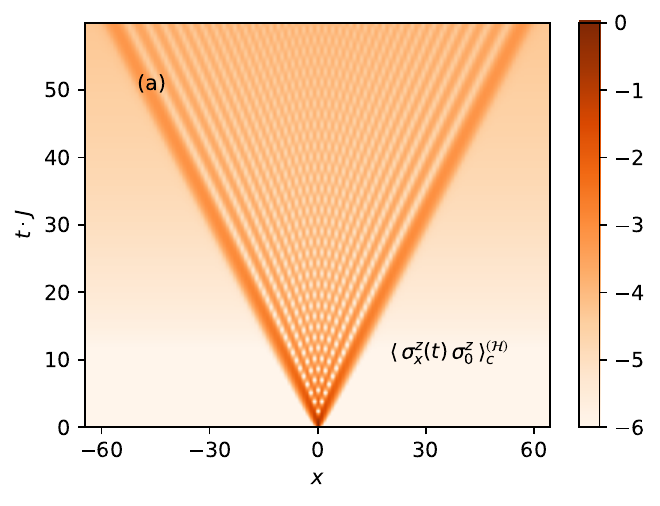}
  \includegraphics[width=0.49\textwidth]{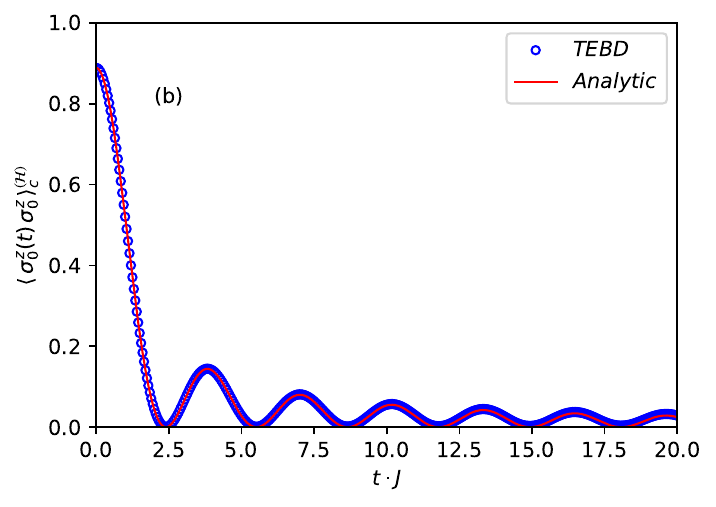}
   \caption{(a) Density plot showing the time evolution of the connected
   longitudinal  correlation function $\average{\sigma_x^{z}(t)\,\sigma_{0}^{z}}^{(\mathcal{H})}_c$ after the quench.
   The data is displayed on a logarithmic scale to enhance the visibility of the ballistic light-cone structure.
   (b) Temporal evolution of the autocorrelation at the central site, comparing TEBD results with the analytical expression Eq.~\eqref{eq:S0_zz}. Simulation parameters: $L=129$, $\zeta=2$.}
  \label{fig:S_zz}
  \end{center}
\end{figure}
In deriving this equation, we have assumed the thermodynamic limit, which allows us to extend the summation 
over $l'$ and $l''$ to infinity, and  allows us to use the identity $\sum_{l'=-\infty}^{\infty}J_{l'}^2(x)=1$ to simplify 
the expression. Equation~\eqref{eq:S0_zz} highlights that despite the complete 
absence of correlations in the initial steady state, nontrivial longitudinal correlations are dynamically generated by the coherent evolution. 
Equivalently, the connected autocorrelation is particularly simple in the thermodynamic limit,
\begin{align}
\average{\sigma_0^{z}(t)\,\sigma_{0}^{z}}^{(\mathcal{H})}_{\!c}
& \equiv
\average{\sigma_0^{z}(t)\,\sigma_{0}^{z}}^{(\mathcal{H})}-\langle \sigma^z \rangle_{\mathrm{SS}}^2\nonumber\\
& =
\big(1-\langle \sigma^z \rangle_{\mathrm{SS}}^2\big)\,J_0^2(Jt),
\label{eq:S0_zz_connected}
\end{align}
showing explicitly that its amplitude is controlled solely by the fugacity $\zeta$ through Eq.~\eqref{eq:Sz_NESS}.
Furthermore, the correlator relaxes to the steady-state value $\langle \sigma^z \rangle_{\mathrm{SS}}^2$, with an algebraic approach governed by $J_0^2(Jt)\sim (\pi J t)^{-1}$ at long times, which is a direct consequence of the ballistic spreading of correlations, associated with a dynamical exponent $z=1$.

The thermodynamic limit result generalizes straightforwardly to arbitrary separation $\ell$.
Using the same free-fermion propagator and Wick’s theorem for the infinite-temperature product state with filling $\bar n$, one finds
\begin{equation}
\average{\sigma_{\ell}^{z}(t)\,\sigma_{0}^{z}}^{(\mathcal{H})}
=
\langle \sigma^z \rangle_{\mathrm{SS}}^2
+\big(1-\langle \sigma^z \rangle_{\mathrm{SS}}^2\big)\,J_{\ell}^{2}(Jt),
\label{eq:Szz_sep_XX}
\end{equation}
and therefore $\average{\sigma_{\ell}^{z}(t)\,\sigma_{0}^{z}}^{(\mathcal{H})}_{\!c}=(1-\langle \sigma^z \rangle_{\mathrm{SS}}^2)\,J_{\ell}^{2}(Jt)$.
Equation~\eqref{eq:Szz_sep_XX} makes the light-cone structure explicit: for fixed $t$ the spectral weight is concentrated at
$|\ell|\lesssim Jt$, while the autocorrelation decays as $J_\ell^2(Jt)\sim (\pi Jt)^{-1}$ at late times.

Figure~\ref{fig:S_zz}(a) illustrates the formation of a light-cone structure in the longitudinal correlation function following 
the quench from the nonequilibrium steady state. The correlations spread ballistically from the initially perturbed site, with 
a well-defined causal front whose slope is set by the maximal group velocity $v\approx J$. Inside the light cone, nontrivial 
correlations build up dynamically, while outside the cone correlations remain strongly suppressed. 

In Fig.~\ref{fig:S_zz}(b), we compare the time dependence of the longitudinal correlation function at the central site obtained 
from TEBD simulations with the analytical expression in Eq.~\eqref{eq:S0_zz} for 
$\zeta=2.0$ and $\Delta=0$ which display a perfect match between numerical data and the analytical result.

\subsection{Interacting case, $\Delta\neq 0$}\label{sec:Delta_nonzero}
\begin{figure*}[tbh!]
  \begin{center}
  \includegraphics[width=0.32\textwidth]{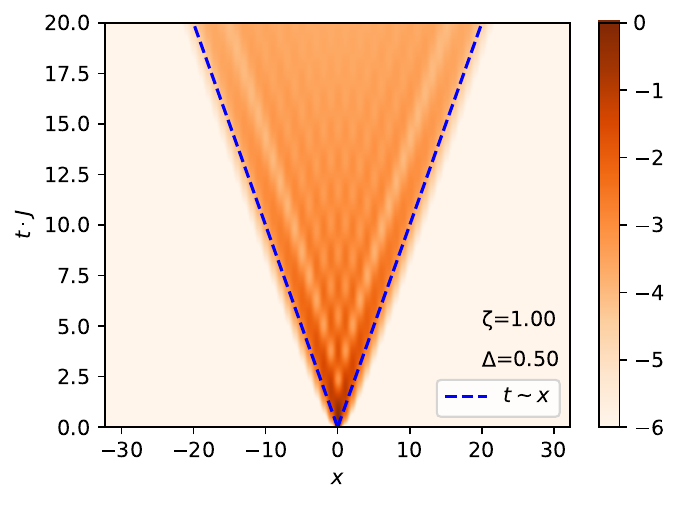}
  \includegraphics[width=0.32\textwidth]{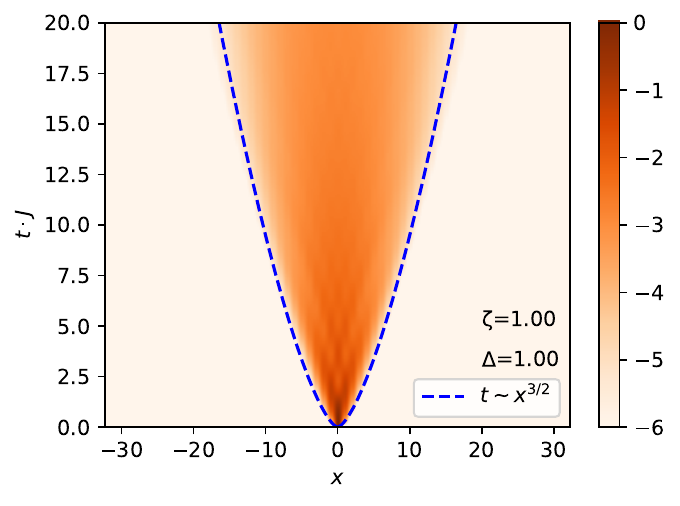}
  \includegraphics[width=0.32\textwidth]{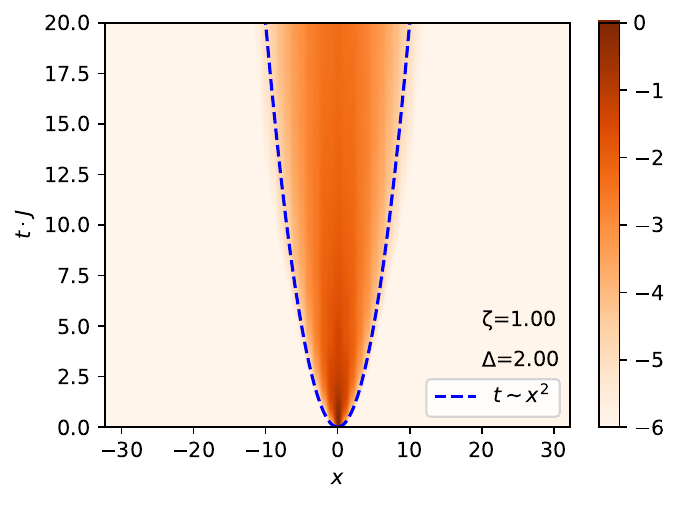}
  \includegraphics[width=0.32\textwidth]{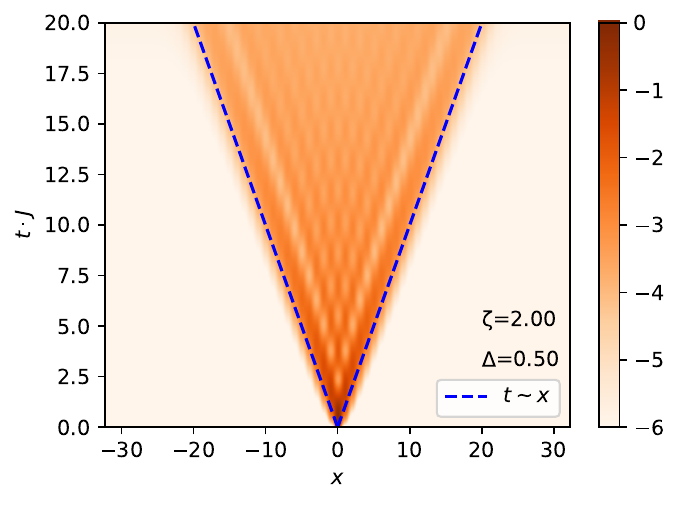}
  \includegraphics[width=0.32\textwidth]{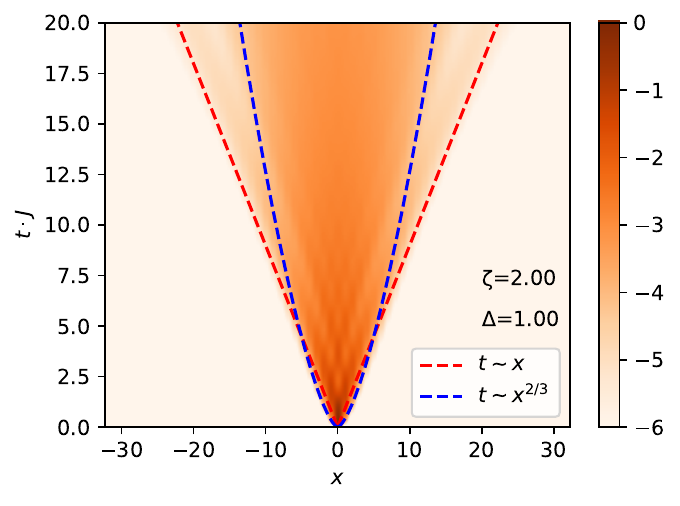}
  \includegraphics[width=0.32\textwidth]{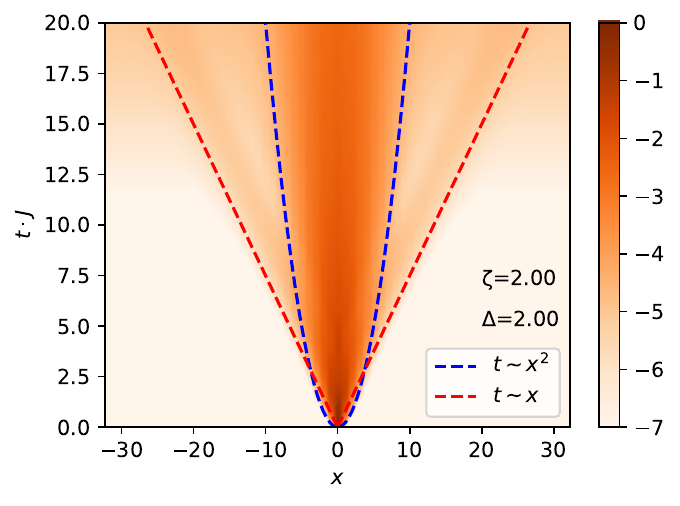}	
   \caption{Time evolution of the longitudinal autocorrelation $\langle \sigma_x^z(t)\,\sigma_0^z \rangle_{c}^{\mathcal{(H)}}$ for
   several values of $\Delta$ and fugacity $\zeta$ (indicated in each panel). The data is displayed on a logarithmic color scale to enhance the visibility of the light-cone structure.
   Dashed lines mark the light-cone boundaries. The dynamical exponent $z$ governing the cone
   is ballistic ($z=1$) for $\Delta<1$, superdiffusive ($z=3/2$) at $\Delta=1$, and diffusive ($z=2$) for $\Delta>1$.
   For $\Delta=2.0$, $\zeta=2.0$ two distinct fronts are visible, signaling coexistence of ballistic and diffusive modes.  Data shown on a logarithmic color scale.}
  \label{fig:S_zz_Delta_panels}
  \end{center}
\end{figure*}

There is no closed analytical expression for $\Delta\neq 0$, and we rely on TEBD.
Figure~\ref{fig:S_zz_Delta_panels} presents the spatiotemporal structure of the longitudinal autocorrelation
across the three transport regimes predicted by GHD~\cite{castroalvaredo2016emergent,bertini2016transport}. 
For $\zeta=1$ (zero magnetization, infinite temperature) the light cone is ballistic for $\Delta<1$, 
superdiffusive at $\Delta=1$~\cite{ljubotina2019kardar,ilievski2018superdiffusion}, and diffusive for $\Delta>1$~\cite{dupont2020universal}.
For $\zeta=2.0$, the non-zero background magnetization influences  the ensuing dynamics by introducing
another channel for the propagation of correlation due to the presence of finite magnon density\cite{weiner2020}, stemming from the finite longitudinal magnetization.
Due to this, at $\Delta=2$, two propagating fronts coexist — an outer ballistic front associated with magnon excitations and an inner 
diffusive one~\cite{znidaric2019coexistence, medenjak2021exact,weiner2020}.

In order to extract the dynamical exponent, we focus on the long-time power-law decay of the connected autocorrelation,
$\average{\sigma_0^{z}(t)\,\sigma_{0}^{z}}^{(\mathcal{H})}_{\!c}$, as defined in Eq.~\eqref{eq:S0_zz_connected}. This quantity probes the relaxation of a local magnetization fluctuation on top of the SS,
which is expected to exhibit a universal power-law decay,
\begin{equation}
\average{\sigma_0^{z}(t)\,\sigma_{0}^{z}}^{(\mathcal{H})}_{\!c}
\propto
(Jt)^{-1/z},
\label{eq:Szz_scaling}
\end{equation}
with $z$ the dynamical exponent.
The exactly solvable $\Delta=0$ benchmark gives
$\average{\sigma_0^{z}(t)\,\sigma_{0}^{z}}_{c}^{(\mathcal{H})}\propto J_0^2(Jt)\sim (\pi J t)^{-1}$,
corresponding to ballistic transport with $z=1$.
\begin{figure}[tbh!]
  \begin{center}
   \includegraphics[width=0.49\textwidth]{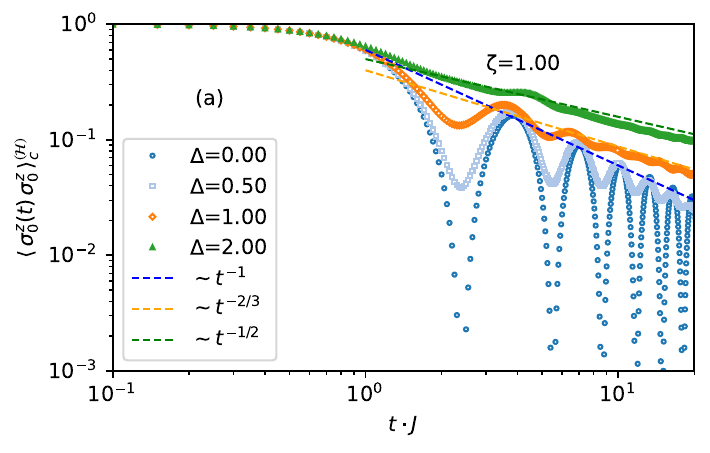}
   \includegraphics[width=0.49\textwidth]{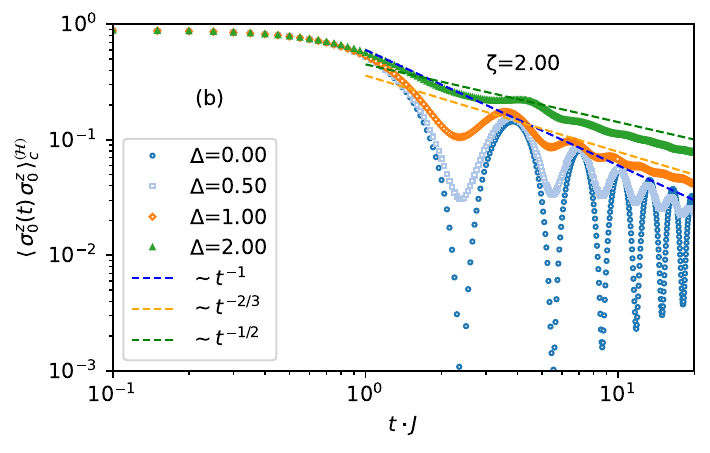}
   \caption{Long-time decay of the central longitudinal autocorrelation for different fugacities $\zeta$.
   The asymptotic behavior is consistent with $\average{\sigma_0^{z}(t)\,\sigma_{0}^{z}}^{(\mathcal{H})}_{\!c}\sim (Jt)^{-1/z}$ irrespective of $\zeta$, confirming that the dynamical exponent $z$ is independent of the initial magnetization and is solely determined by the anisotropy $\Delta$.}
  \label{fig:S_zz_Delta_rescaled}
  \end{center}
\end{figure}
Figure~\ref{fig:S_zz_Delta_rescaled} reveal two effects of interactions:
 {\it (i)} Interactions generate pronounced short-time oscillations and a fast initial decay;
these are nonuniversal and depend on $\zeta$ through the background magnetization $\langle\sigma^z\rangle_{\mathrm{SS}}$.
{\it (ii)} At late times, the decay crosses over to universal power-law relaxation~\cite{bertini2021finite,gopalakrishnan2018hydrodynamics}.
In the easy-plane phase ($\Delta<1$), we find $z\approx 1$ (ballistic), continuously connected to the $\Delta=0$ result.
At the isotropic point ($\Delta=1$), the data are consistent with KPZ superdiffusion, $z\approx 3/2$~\cite{ljubotina2017spin, bulchandani2021kardar}.
In the easy-axis regime ($\Delta>1$), diffusive decay $z\approx 2$ is observed~\cite{dupont2020universal,medenjak2021exact}.
These three cases translate to $\average{\sigma_0^{z}(t)\,\sigma_{0}^{z}}^{(\mathcal{H})}_{\!c}\propto t^{-1}$, $t^{-2/3}$, and $t^{-1/2}$, respectively.

Importantly, the dynamical exponent $z$ appears to be insensitive to $\zeta$ within our numerical accuracy, confirming its universal character.
Interestingly, for $\zeta=2$ (finite longitudinal magnetization, away from infinite temperature), although the spatiotemporal light cone at $\Delta=1$ does not display the 
characteristic KPZ spreading pattern, the power-law decay of the autocorrelation at the central site is still consistent with the KPZ exponent $t^{-2/3}$.
We suspect that this is a results of the intricate interplay between KPZ and ballistic propagation, visible in Fig.~\ref{fig:S_zz_Delta_panels}. 
The central site decay is expected to be a superposition  of ballistic, $1/t$ and KPZ, $t^{-2/3}$ decay profiles, out of which the KPZ wins and dominates the temporal decay after some early time
transient.
This suggests that the KPZ universality class controls the long-time decay of the autocorrelator even away from the infinite-temperature limit, a finding that 
goes beyond the standard setting in which KPZ scaling has been established~\cite{ljubotina2019kardar, ilievski2018superdiffusion}.

Identical conclusions apply to the $\Delta>1$, diffusive case as well, where a coexistence of ballistic, $1/t$ and diffusive $t^{-1/2}$ decay is expected
for the local correlator from the light cone structures. Among these, the diffusive decay dominates over the subleading ballistic decay.

\section{Unitary evolution, Transverse correlations}\label{sec:transverse_unitary}
The transverse spin-spin correlator $\average{\sigma_l^{x}(t)\,\sigma_{l'}^{x}}^{(\mathcal{H})}$ probes 
spin coherence: it involves an operator that does not commute with the XXZ Hamiltonian and hence carries 
information about both spin transport and quantum information spreading~\cite{sachdev2000quantum, mikeska2004one}. 

In the noninteracting limit $\Delta=0$, the transverse correlator can be expressed in terms of 
hard-core bosonic correlators, which admit a determinant representation via summation over form 
factors~\cite{KBI93,CIKT93,SGPM21,Wang22,Patu22}, but also it allows for a direct solution via the 
Jordan--Wigner mapping to free fermions. 
We will present both approaches, which yield identical results. 

\subsection{Noninteracting limit, $\Delta=0$: Determinant representation}\label{sec:transverse_Delta_0}

At $\Delta=0$, the XXZ chain reduces to the XX model, which can be written in terms of hard-core bosons as
\begin{equation}\label{XXbosonic}
H_{XX}=\frac{J}{2}\sum_{l=-L/2}^{L/2-1}\left(b_l^\dagger b_{l+1}+ \mathrm{h.c.}\right)\, .
\end{equation}
Here $b_l^\dagger$ and $b_l$ are hard-core bosonic operators. On different sites they obey the usual bosonic commutation relations, 
while the on-site constraint $(b_l^\dagger)^2=(b_l)^2=0$ forbids double occupancy. The mapping from spin-$1/2$ operators is
\begin{equation}
  \frac{1}{2}\sigma_l^+ \to b_l^\dagger, \quad \frac{1}{2}\sigma_l^- \to b_l, \quad \sigma_l^z \to 2 b_l^\dagger b_l - 1.
\end{equation}
In this representation, the transverse spin-spin correlator naturally splits into two hard-core-boson correlators,
\begin{equation}\label{eq:sxxbosons}
\langle \sigma_l^x(t) \sigma_{l'}^x \rangle^{(\mathcal{H})}
=
\langle b_l^\dagger(t)\, b_{l'}\rangle^{(\mathcal{H})}
+
\langle b_l(t)\, b_{l'}^\dagger\rangle^{(\mathcal{H})}\, ,
\end{equation}
where
\begin{equation}
\langle b_l^\dagger(t)\, b_{l'}\rangle^{(\mathcal{H})} = \mathrm{Tr}\!\left\{ b_l^\dagger(t)\, b_{l'}\, \rho_{\mathrm{SS}} \right\},
\quad \langle b_l(t)\, b_{l'}^\dagger\rangle^{(\mathcal{H})} = \mathrm{Tr}\!\left\{ b_l(t)\, b_{l'}^\dagger\, \rho_{\mathrm{SS}} \right\}
\end{equation}
while in Eq.~\eqref{eq:rhoSS}, $\mathds{1}_M$ denotes the identity operator in the subspace with $M$ hard-core bosons. The 
advantage of this decomposition is that the full many-body correlator can be reconstructed from a small set of 
single-particle building blocks.

An exact determinant representation of these correlators follows from a form-factor expansion~\cite{KBI93,CIKT93,SGPM21,Wang22,Patu22}. 
For open boundary conditions, the single-particle eigenfunctions and eigenenergies of Eq.~\eqref{XXbosonic} are 
\begin{equation}
\phi_n(l)=\left(\frac{2}{L+1}\right)^{1/2}\sin\!\left(\frac{n \pi l}{L+1}\right),
\quad \varepsilon_n=J \cos\!\left(\frac{\pi n}{L+1}\right)\nonumber
\end{equation}
with time evolution $\phi_n(l,t)=e^{-i \varepsilon_n t}\phi_n(l)$. The basic propagation kernel is 
$g(l,l';t)=\sum_{k=1}^L \phi_k(l,t)\,\overline{\phi_k(l')}$, while the nonlocal string structure is encoded in the auxiliary function
$f(k,q\,|\,l,t)=\delta_{k,q}-2\sum_{z=l}^{L} \phi_k(z,t)\,\overline{\phi_q(z,t)}$, where the overbar denotes complex 
conjugation. In terms of these ingredients, we define the matrices $U^{(\pm)}$ and $R^{(\pm)}$ with  
the components ($a,b=1,\dots,L$)
\begin{align}
U_{a,b}^{(-)}(l,l'; t)&= \sum_{q=1}^L \overline{f(a,q\,|\, l,t)}\, f(b,q\,|\,l',0), \\
U_{a,b}^{(+)}(l,l'; t)&= \sum_{k=1}^L f(k,b\,|\, l,t)\, \overline{f(k,a\,|\, l',0)}, 
\end{align}
\begin{align}
R_{a,b}^{(-)}(l,l'; t)&= \overline{\phi_a(l,t)}\, \phi_b(l'), \quad
R_{a,b}^{(+)}(l,l'; t)= \overline{e_a(l,l'; t)}\, e_b(l,l'; t),
\end{align}
where the vectors $e(l,l';t)$ and $\overline{e}(l,l';t)$ have components $e_a(l,l';t)=\sum_{k=1}^L f(k,a\,|\, l,t)\,
\overline{\phi_k(l')}$ and $\overline{e}_a(l,l';t)=\sum_{k=1}^L \overline{f(k,a\,|\, l',0)}\, \phi_k(l,t)$. This 
organization is useful because $U^{(\pm)}$ captures the string-dressed propagation through the chain, whereas $R^{(\pm)}$ 
carries the extra local contribution associated with the endpoint operators.
With this notation, the two bosonic correlators take the determinant form
%
\begin{align}
\langle b_l(t)\, b_{l'}^\dagger\rangle^{(\mathcal{H})}
&=
\det_L\!\left(\frac{\mathbf{1}}{1+\xi}+\frac{\xi}{1+\xi}\left[U^{(-)}+R^{(-)}\right]\right)
 -\det_L\!\left(\frac{\mathbf{1}}{1+\xi}+\frac{\xi}{1+\xi} U^{(-)}\right), \label{eq:gminus}\\
\langle b_l^\dagger(t)\, b_{l'}\rangle^{(\mathcal{H})}
&=
\det_L\!\left(\frac{\mathbf{1}}{1+\xi}+\frac{\xi}{1+\xi}\left[U^{(+)}-R^{(+)}\right]\right)\nonumber \\
&\qquad\qquad\qquad\qquad\qquad+\bigl[g(l,l';t)-1\bigr]\,
\det_L\!\left(\frac{\mathbf{1}}{1+\xi}+\frac{\xi}{1+\xi} U^{(+)}\right), \label{eq:gplus}
\end{align}
%
where $\mathbf{1}$ is the $L\times L$ identity matrix. The two determinants differ only through the 
sign structure and the rank-one terms built from $R^{(\pm)}$, making the relation between particle and 
hole propagation explicit. Together with Eq.~\eqref{eq:sxxbosons}, Eqs.~\eqref{eq:gminus} and \eqref
{eq:gplus} provide an exact and numerically efficient evaluation of the transverse correlator for 
systems of up to several hundred sites.
\begin{figure*}[tbh!]
  \begin{center}
  \includegraphics[width=0.48\textwidth]{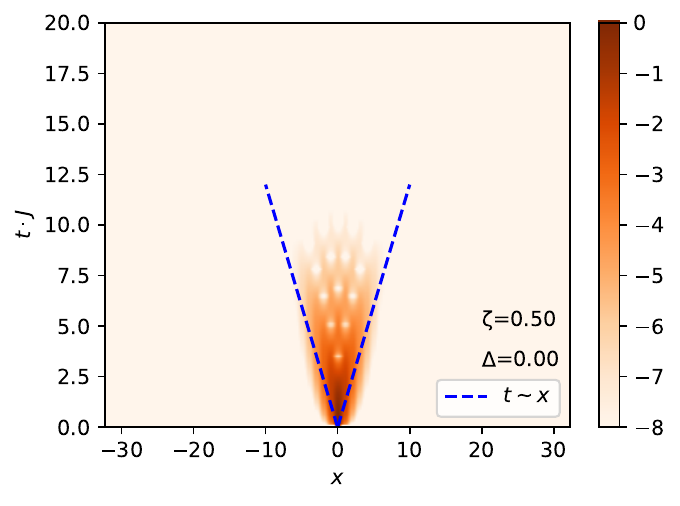}
  \includegraphics[width=0.48\textwidth]{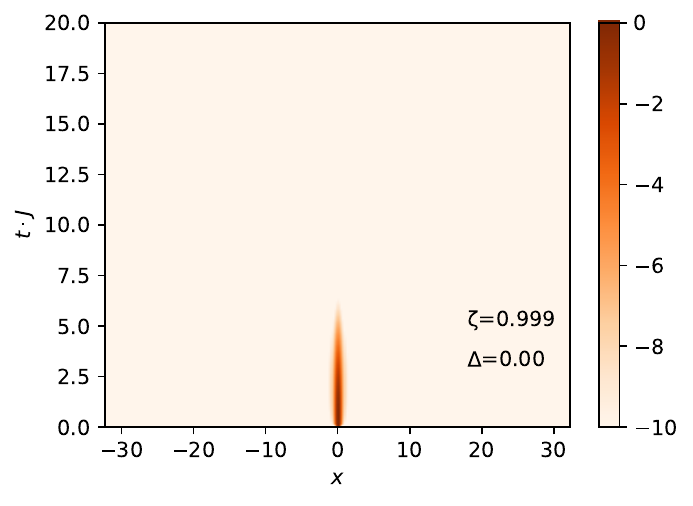}
  \caption{
  Time evolution of the transverse correlator $\average{\sigma_l^{x}(t)\,\sigma_{0}^{x}}^{(\mathcal{H})}$ for $\Delta = 0$ and  $\zeta=\{0.5, 1.0\}$.
  The data is displayed on a logarithmic color scale to enhance the visibility of the spatio-temporal structure. For $\zeta \approx 1$, the spectral function is localized at the center and shows no light-cone structure.}
  \label{fig:S_xx_Delta_0_panels}
  \end{center}
\end{figure*}
In Fig.~\ref{fig:S_xx_Delta_0_panels}, we show the spatiotemporal structure of the transverse correlation function for several values of 
$\zeta$, computed using the determinant approach. For $\zeta \approx 1$, the spectral function is localized at the center and exhibits no 
light-cone structure. 
For $\zeta \neq 1$, an exponential envelope emerges, as indicated by the blue dashed line, which 
is a direct consequence of the finite magnetization in the initial state. The TEBD data (not shown here)perfectly match the results obtained from the determinant approach.

\subsection{Noninteracting limit, $\Delta=0$: Pfaffian approach}\label{sec:transverse_Delta_0_pfaffian}

In the noninteracting limit $\Delta=0$, the  calculation can also be reduced to a free-fermion 
problem via the Jordan--Wigner transformation~\cite{lieb1961two, katsura1962statistical}. 
The transverse spin operator can be written as a nonlocal fermionic string,
\begin{equation}
\sigma_l^{x}
=
\exp\!\left(
i\pi\sum_{l'<l} c_{l'}^{\dagger} c_{l'}
\right)
\left(
c_l + c_l^{\dagger}
\right).
\end{equation}
Employing the identity $(-1)^{c_l^{\dagger}c_l}=1-2c_l^{\dagger}c_l=(c_l^{\dagger}+c_l)(c_l^{\dagger}-c_l)$ and introducing Majorana operators
\begin{equation}
\eta_{2l-1}=c_l+c_l^{\dagger},
\qquad
\eta_{2l}=-i\left(c_l-c_l^{\dagger}\right),
\end{equation}
which satisfy $\{\eta_a,\eta_b\}=2\delta_{ab}$, the Jordan--Wigner string is rewritten in a compact local form,
\begin{equation}
(-1)^{c_l^{\dagger}c_l}
=
-i\,\eta_{2l-1}\eta_{2l}.
\end{equation}
As a result, the transverse spin operator at the origin takes the form
\begin{equation}
\sigma_0^{x}
=
\prod_{l<0}
\left(
-i\,\eta_{2l-1}\eta_{2l}
\right)
\eta_{-1}.
\end{equation}

\begin{figure}[tbh!]
  \begin{center}
   \includegraphics[width=0.6\textwidth]{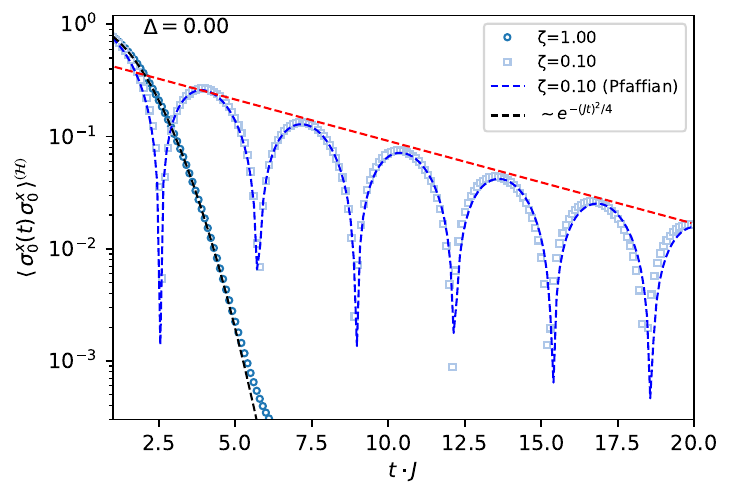}
   \caption{Transverse autocorrelation $\average{\sigma_0^{x}(t)\,\sigma_0^{x}}^{(\mathcal{H})}$ at $\Delta=0$ 
   for several fugacities $\zeta$. Symbols represent the TEBD data while the  dashed lines represent the exact Pfaffian result. At 
   $\zeta=1$ the decay is Gaussian, Eq.~\eqref{eq:S_xx_Delta_0_z_1}; for $\zeta\neq 1$ an exponential envelope emerges as indicated by the red dashed line.}
  \label{fig:S_xx_Delta_0}
  \end{center}
\end{figure}

\begin{figure*}[tbh!]
  \begin{center}
  \includegraphics[width=0.32\textwidth]{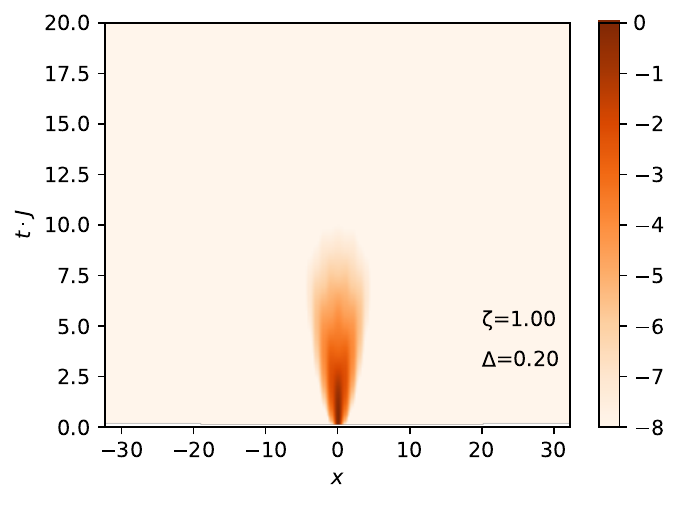}
  \includegraphics[width=0.32\textwidth]{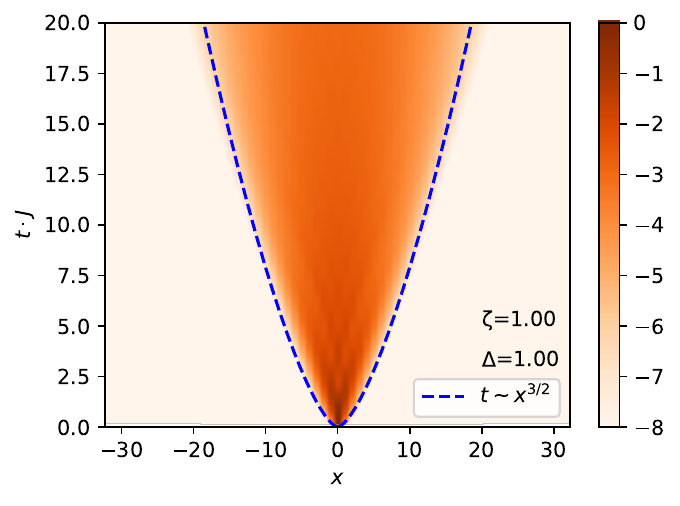}
  \includegraphics[width=0.32\textwidth]{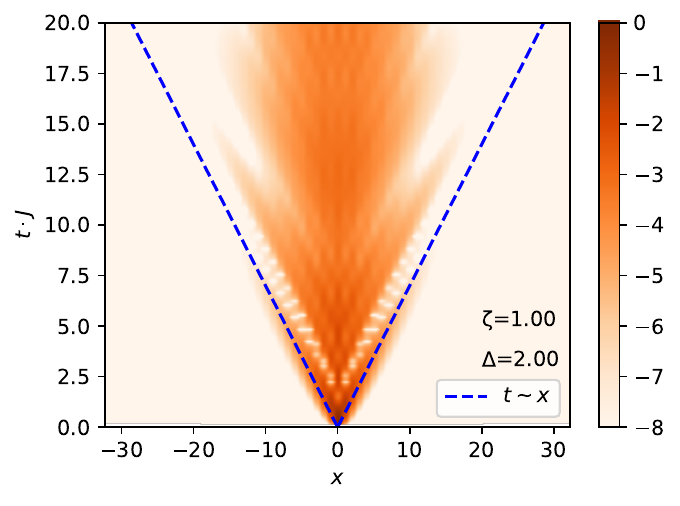}
  \includegraphics[width=0.32\textwidth]{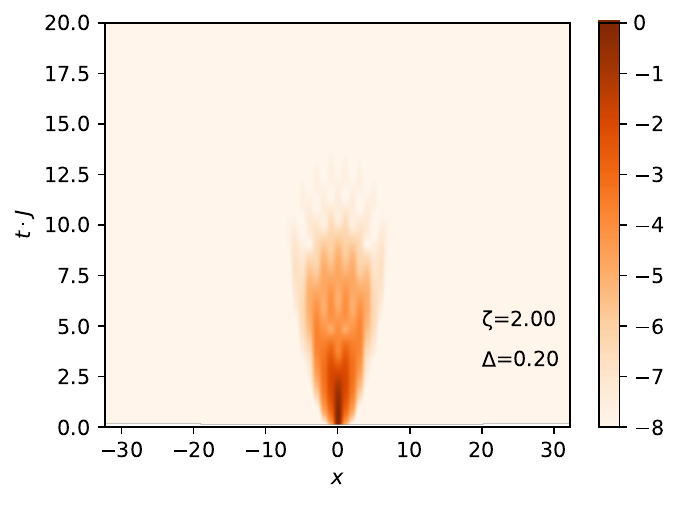}
  \includegraphics[width=0.32\textwidth]{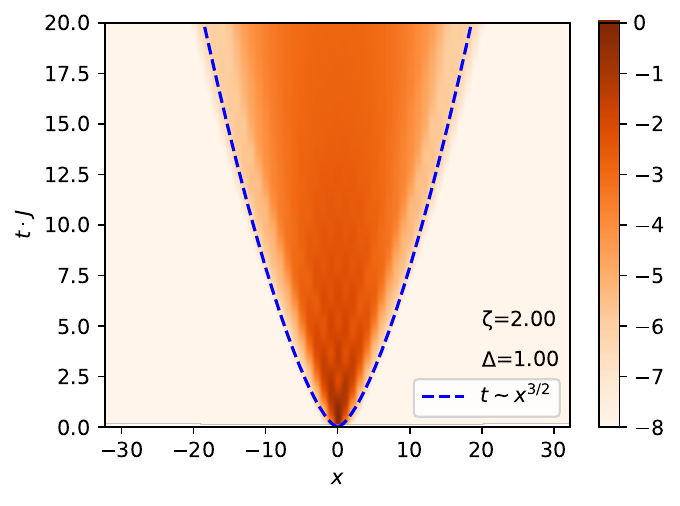}
  \includegraphics[width=0.32\textwidth]{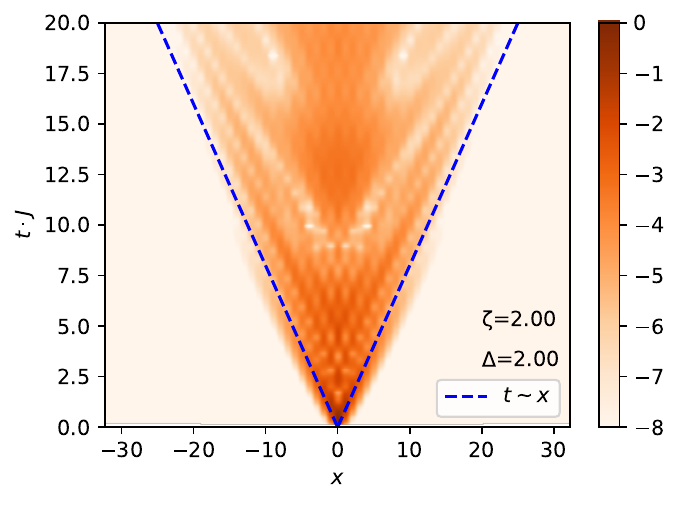}	
  \caption{Time evolution of the transverse correlator $\average{\sigma_l^{x}(t)\,\sigma_{0}^{x}}^{(\mathcal{H})}$ for
  several values of $\Delta$ and fugacity $\zeta$ (indicated in each panel). The data is displayed on a logarithmic color scale to enhance the visibility of the spatiotemporal structure.
  Dashed lines mark the front positions where a propagating wavefront is visible. At $\Delta=1$ the front is superdiffusive ($z=3/2$), while for $\Delta>1$ it is ballistic ($z=1$).
  For $\Delta<1$, the signal remains localized near the origin and decays rapidly without a propagating light cone.}
  \label{fig:S_xx_Delta_panels}
  \end{center}
\end{figure*}

\begin{figure}[tbh!]
  \begin{center}
   \includegraphics[width=0.55\textwidth]{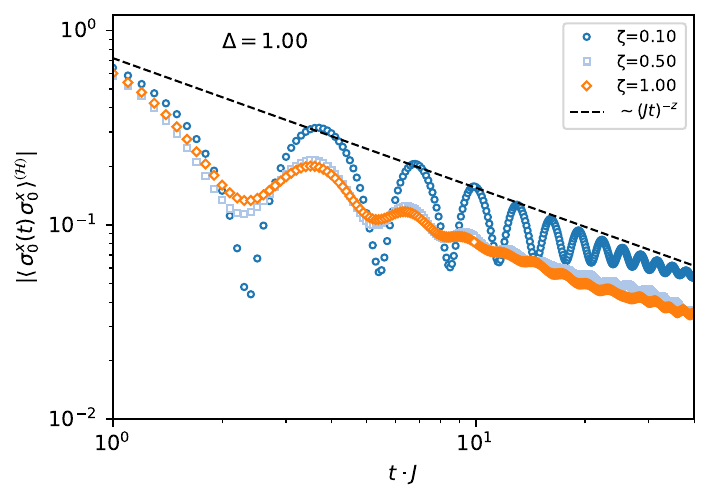}
   \caption{Transverse autocorrelation at the isotropic point ($\Delta=1$) for several fugacities $\zeta$. The long-time power-law decay is consistent with KPZ superdiffusion ($z=3/2$) for all $\zeta$, demonstrating that KPZ universality persists away from infinite temperature.}
  \label{fig:S_xx_Delta_1}
  \end{center}
\end{figure}

\begin{figure}[tbh!]
  \begin{center}
   \includegraphics[width=0.6\columnwidth]{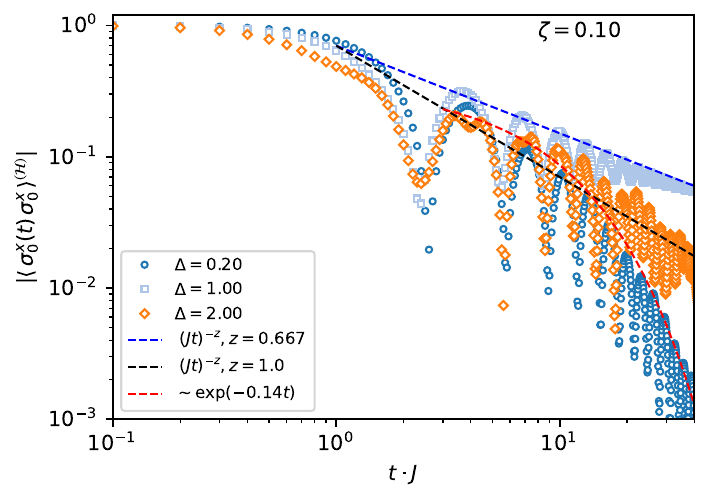}
   \caption{Transverse autocorrelation vs.\ anisotropy $\Delta$ at fixed $\zeta=0.1$. The decay crosses over from exponential for $\Delta<1$, through KPZ at $\Delta=1$, to an algebraic tail $\sim(Jt)^{-1/z}$ with $z=1$ for $\Delta>1$.}
  \label{fig:S_xx_Delta_scan}
  \end{center}
\end{figure}

The transverse spin-spin correlation function under unitary evolution can therefore be expressed as
\begin{eqnarray}
\average{\sigma_0^{x}(t)\,\sigma_{0}^{x}}^{(\mathcal{H})}
&=&
(-1)^{(L-1)/2}
\Big\langle
\Big[
\prod_{l<0}
\eta_{2l-1}(t)\eta_{2l}(t)
\Big]
\eta_{-1}(t)
\Big[
\prod_{l<0}
\eta_{2l-1}\eta_{2l}
\Big]
\eta_{-1}
\Big\rangle .
\label{eq:S_xx}
\end{eqnarray}
The operator string in Eq.~\eqref{eq:S_xx} contains an even number of Majorana operators.
Since the XX dynamics is quadratic, the time-evolved state remains Gaussian~\cite{bravyi2004lagrangian, peschel2003calculation} and all multi-point Majorana correlators factorize via Wick's theorem.
The correlator therefore reduces to a Pfaffian~\cite{mccoy1971spin, franchini2017introduction},
\begin{equation}
\average{\sigma_0^{x}(t)\,\sigma_{0}^{x}}^{(\mathcal{H})}
=
\operatorname{Pf}
\!\left(
\Gamma^{(\sigma^x)}
\right),
\end{equation}
where $\Gamma^{(\sigma^x)}$ is the antisymmetric matrix of two-point Majorana averages for the operator string in Eq.~\eqref{eq:S_xx}; its entries are listed in Appendix~\ref{sec:Majorana}.
The Pfaffian can be evaluated numerically for any $\zeta$. At $\zeta=1$ (half-filling), the spectral flatness of the XX dispersion at the Fermi level leads to a Gaussian autocorrelation~\cite{Sur1975,Brandt1976,bunder1999effect},
\begin{equation}
\average{\sigma_0^{x}(t)\,\sigma_{0}^{x}}^{(\mathcal{H})} \approx e^{-{1\over 4} J^2 t^2}.
\label{eq:S_xx_Delta_0_z_1}
\end{equation}
Away from $\zeta=1$, the Gaussian form changes, and an exponential envelope with a $\zeta$-dependent rate emerges, overlaid on bandwidth 
oscillations. 
Figure~\ref{fig:S_xx_Delta_0} confirms excellent quantitative agreement between TEBD and the exact Pfaffian result for all 
fugacities.

\subsection{Interacting case, $\Delta\ne 0$}\label{sec:transverse_Delta_ne_0}
For interacting $\Delta\neq 0$, no Pfaffian representation is available and we rely on TEBD~\cite{paeckel2019time, vidal2007classical}.
The full space-time structure of $\average{\sigma_l^{x}(t)\,\sigma_{0}^{x}}^{(\mathcal{H})}$ is summarized in Fig.~\ref{fig:S_xx_Delta_panels}.
The transverse correlations are largely unaffected by the possible finite longitudinal magnetization as this affects a distinct spin component. Three qualitatively distinct regimes are apparent:

\emph{(i)} In the easy-plane phase ($\Delta<1$) the transverse spectrum localizes the initial perturbation: the signal decays rapidly on a 
few lattice sites with a near-Gaussian spatial profile, and no genuine wavefront propagates into the bulk~\cite{sachdev2000quantum, giamarchi2003quantum}.

\emph{(ii)} At the isotropic point ($\Delta=1$) a propagating front emerges whose position scales as $x\sim t^{2/3}$, consistent with the KPZ dynamical 
exponent $z=3/2$ inherited from the $SU(2)$-symmetric transport~\cite{ljubotina2019kardar, ilievski2018superdiffusion, krajnik2020kardar}.
The $SU(2)$ symmetry of the Heisenberg chain~\cite{takahashi1999thermodynamics} ties the transverse and longitudinal sectors together, so the transverse 
auto-correlation inherits the KPZ superdiffusion established for the longitudinal channel~\cite{ljubotina2019kardar, ilievski2018superdiffusion, bulchandani2021kardar}. 
Figure~\ref{fig:S_xx_Delta_1} confirms the universal $t^{-2/3}$ decay for all simulated fugacities $\zeta$, demonstrating that KPZ universality at $\Delta=1$ persists away from infinite temperature and is controlled by the $SU(2)$ symmetry rather than by a specific filling~\cite{krajnik2020kardar}.

\emph{(iii)} In the easy-axis regime ($\Delta>1$) a sharp ballistic light cone $x\sim t$ is formed. 
This is the typical behaviour expected for the dynamics of a non-conserved quantity and signals the presence of long-lived magnon excitations.

The apparent difference between the $\Delta<1$ and $\Delta>1$ behaviours can be understood in the fermionic language by focusing of the role of string operators.
The Jordan-Wigner string depends on $\sigma^z$ only, while the $\Delta<1$ Hamiltonian, governing the time evolution, depends dominantly on $\sigma^{x,y}$. Their interplay
leads to very strong fluctuations in the string operator, eventually destroying the propagation of correlations. As opposed to that, in the $\Delta>1$ limit,
both the strings and the time evolving Hamiltonian depend dominantly on $\sigma^z$, therefore the string operator becomes  non-destructive for correlation propagation.

The anisotropy dependence at fixed $\zeta=0.1$ is shown in Fig.~\ref{fig:S_xx_Delta_scan}.
In the easy-plane phase ($\Delta<1$), the transverse correlator decays exponentially, reflecting the absence of diffusion in the transverse channel 
for a $U(1)$-symmetric gapless system~\cite{giamarchi2003quantum}.
At $\Delta=1$ the KPZ power law is recovered.
In the easy-axis regime ($\Delta>1$), the transverse amplitude decays ballistically: the wavefront propagates at the magnon group velocity and the 
envelope follows a $\sim(Jt)^{-1}$ power law ($z=1$), reflecting the coherent propagation of magnons rather than diffusive 
spreading~\cite{gopalakrishnan2018hydrodynamics, znidaric2011spin,weiner2020} in the transverse channel.

\section{Lindbladian evolution, Longitudinal correlations}\label{sec:Lindblad_longitudinal}

We now turn on full Lindbladian dynamics, where the coherent XXZ exchange competes with local incoherent spin-flip processes.
The open dynamics is governed by the GKSL master equation~\cite{lindblad1976generators, gorini1976completely, breuer2002theory},
which drives the system toward the SS at fugacity $\zeta$.
Throughout this section the system is initialised in the SS at the same fugacity $\zeta$, so that the background filling is stationary and changes in the correlator arise purely from the coherent and dissipative propagation of the perturbation inserted by the second spin operator. The two essential ingredients of the Lindbladian dynamics are: 

\emph{(i)} coherent XXZ exchange, which spreads
magnetization fluctuations (ballistically, KPZ or diffusively) and generates a light-cone 
structure~\cite{lieb1972finite, bravyi2006lieb} 

\emph{(ii)} local incoherent spin-flip processes  which act as a spatially homogeneous source of decoherence. These two 
contributions define a crossover time scale $t^{*(z)}_{\rm eff}$~\cite{damanet2019atom, diehl2008quantum, verstraete2009quantum}: for short times $t\ll t^{*(z)}_{\rm eff}$ 
coherent exchange dominates, the dissipative envelope is negligible, and the light cone builds up as in 
the purely unitary problem; the \emph{shape} of the causal front — and hence the universality class — 
is set during this regime. For $t\gtrsim t^{*(z)}_{\rm eff}$ exponential damping takes over and 
suppresses all correlations uniformly, so the light cone loses its  meaning even though the 
causal velocity itself is unaltered. The effective rate $\Gamma_{\rm eff}^{(z)}$ 
reduces to the bare $\Gamma=\gamma_l+\gamma_p$ in the noninteracting limit $\Delta=0$, but acquires  a $\Delta$ and $\zeta$-dependence  in the presence of interactions.

\subsection{Noninteracting limit, $\Delta=0$: Exact analytical solution}\label{sec:Lindblad_Delta_0}

In the noninteracting limit, $\Delta=0$, 
we provide an exact analytical expression for the longitudinal 
autocorrelation,  benchmarked against the third-quantization ($3^{\mathrm{rd}}$QT) approach~\cite{prosen2008third, prosen2011open}. 
For $\Delta=0$, the Hamiltonian is quadratic in Jordan--Wigner fermions, Eq.~\eqref{eq:H_spinless}, and the local spin-flip jump operators are linear in fermionic operators~\cite{eisler2011crossover, eisler2014entanglement}.
In the fermionic representation the GKSL equation takes the standard form
\begin{align}
\dot\rho
& =
-i[H_{\mathrm{XX}},\rho]
+\sum_x \gamma_l\Big(c_x\rho c_x^{\dagger}-\tfrac{1}{2}\{c_x^{\dagger}c_x,\rho\}\Big) \nonumber\\
& +\sum_x \gamma_p\Big(c_x^{\dagger}\rho c_x-\tfrac{1}{2}\{c_x c_x^{\dagger},\rho\}\Big),
\label{eq:lindblad_XX_fermions}
\end{align}
The equation of motion for any observable $O$ can be obtained from the adjoint Lindbladian,
\begin{equation}
\dot O
= i[H_{\mathrm{XX}},O]
+\sum_x\Big(L_x^{\dagger} O L_x -\tfrac{1}{2}\{L_x^{\dagger}L_x,O\}\Big),
\end{equation}
with $L_x^- = \sqrt{\gamma_l}\,c_x$ and $L_x^+ = \sqrt{\gamma_p}\,c_x^{\dagger}$. 
When applied to the annihilation and creation operators, this equation of motion closes.  
As a result, one finds a closed equation of motion for the annihilation operators,
\begin{align}
\frac{d}{dt}c_x(t)
&=
-i\frac{J}{2}\big(c_{x+1}(t)+c_{x-1}(t)\big)
-\frac{\Gamma}{2}\,c_x(t),
\label{eq:cl_eom}
\end{align}
where $\Gamma \equiv \gamma_l+\gamma_p$ is the total single-site incoherent rate.
Diagonalizing by Fourier transform, $c_k = L^{-1/2}\sum_x e^{-ikx}c_x$, yields the solution
\begin{equation}
c_k(t)=e^{-\frac{\Gamma}{2}t}\,e^{-i\epsilon(k)t}\,c_k,
\qquad \epsilon(k)=J\cos k,
\label{eq:ck_solution}
\end{equation}
and in real space this becomes the unitary Bessel-function propagator dressed by an overall exponential damping,
\begin{equation}
c_x(t)
=
e^{-\frac{\Gamma}{2}t}
\sum_{x'}
i^{\,x'-x}\,J_{x'-x}(Jt)\,c_{x'}.
\label{eq:cl_solution_bessel_damped}
\end{equation}
The local occupation $n_x(t)=\langle c_x^{\dagger}(t)c_x(t)\rangle$ obeys a continuity equation with local source and sink terms,
\begin{equation}
\frac{d}{dt}n_x(t)
= -\Gamma\,n_x(t) + \gamma_p
 -\big(j_{x}(t)-j_{x-1}(t)\big),
\label{eq:nx_continuity}
\end{equation}
where $j_x(t)=\frac{J}{2i}\langle c_{x+1}^{\dagger}(t)c_x(t)-c_x^{\dagger}(t)c_{x+1}(t)\rangle$ is the coherent particle current.
In a translation-invariant setting, the current divergence vanishes and the occupation relaxes exponentially,
\begin{equation}
n(t)=n_{\mathrm{SS}}+\big(\bar{n}-n_{\mathrm{SS}}\big)e^{-\Gamma t},
\qquad
n_{\mathrm{SS}}=\frac{\gamma_p}{\Gamma}=\frac{\zeta}{1+\zeta},
\label{eq:n_relaxation}
\end{equation}
recovering the steady-state filling associated with the fugacity $\zeta$.
Since we initialise the system in its own SS at fugacity $\zeta$, the background filling is already stationary: $\bar{n}=n_{\mathrm{SS}}$ so $n(t)=n_{\mathrm{SS}}$ for all $t$.
Equation~\eqref{eq:cl_solution_bessel_damped} then makes explicit how the coherent ballistic propagator is preserved at short times, while dissipation produces an overall amplitude decay on the time scale $\Gamma^{-1}$.
For the stationary-background SS, the unequal-time single-particle correlators
follow from Eq.~\eqref{eq:cl_solution_bessel_damped} together with the quantum regression theorem~\cite{lax1963formal, carmichael1993open},
\begin{align}
\average{c_0^{\dagger}(t)c_0}
&= \bar{n}\,e^{-\frac{\Gamma}{2}t}\,J_0(Jt),
\quad
\average{c_0(t)c_0^{\dagger}}
= (1-\bar{n})\,e^{-\frac{\Gamma}{2}t}\,J_0(Jt).
\end{align}
Using Wick's theorem for the Gaussian steady state, one obtains for $t>0$
\begin{equation}
\average{n_0(t)n_0}^{(\mathcal{L})}
=
\bar{n}^2+\bar{n}(1-\bar{n})\,e^{-\Gamma t}J_0^2(Jt).
\end{equation}
Then, the longitudinal spin autocorrelation at the same site reads
\begin{align}
\average{\sigma_0^{z}(t)\,\sigma_{0}^{z}}^{(\mathcal{L})}_c
&=
\average{\sigma_0^{z}(t)\,\sigma_{0}^{z}}^{(\mathcal{L})} -
\langle \sigma^z \rangle_{\mathrm{SS}}^2
=\big(1-\langle \sigma^z \rangle_{\mathrm{SS}}^2\big)\,e^{-\Gamma t}J_0^2(Jt),
\label{eq:S0zz_Lindblad_XX}
\end{align}
which reduces to the unitary result Eq.~\eqref{eq:S0_zz} in the limit $\Gamma\to 0$.
The thermodynamic-limit expression again extends immediately to arbitrary separation $\ell$. Using the 
damped propagator in Eq.~\eqref{eq:cl_solution_bessel_damped}
 together with Wick's theorem for the Gaussian SS, one obtains
\begin{equation}
\average{\sigma_{\ell}^{z}(t)\,\sigma_{0}^{z}}^{(\mathcal{L})}_{\!c}
=
\big(1-\langle \sigma^z \rangle_{\mathrm{SS}}^2\big)\,e^{-\Gamma t}J_{\ell}^{2}(Jt).
\label{eq:Szz_Lindblad_XX}
\end{equation}
This is the direct dissipative analogue of Eq.~\eqref{eq:Szz_sep_XX}: the ballistic XX light cone 
remains encoded in the Bessel profile $J_{\ell}^{2}(Jt)$, with spectral weight concentrated around $|
\ell|\lesssim Jt$, while the bath contributes only the multiplicative damping factor $e^{-\Gamma t}$. 
In this free-fermion limit, dissipation therefore suppresses the amplitude uniformly in time without 
modifying the underlying causal structure set by the coherent hopping.

Equation~\eqref{eq:Szz_Lindblad_XX} also identifies the (noninteracting limit) 
crossover time  as $t^{*} \sim \Gamma^{-1}$. 
For $t\ll t^{*}$, the damping factor remains close to unity, $e^{-\Gamma t}\approx 1-\Gamma t$, 
and the evolution is effectively indistinguishable from the unitary case: the ballistic front forms and 
the light cone is well defined. For $t\gtrsim t^{*}$, however, the exponential decay becomes 
quantitatively important and suppresses the correlator at all spatial separations. 
In this regime the light cone loses its practical meaning, since the coherent front is still present in principle but its weight 
is overwhelmed by the global dissipative decay.

\subsection{Noninteracting limit, $\Delta=0$: Third-quantization approach}\label{sec:Lindblad_Delta_0_3rdQT}
As an independent approach, also applicable in the noninteracting limit, we employ the 
third-quantization ($3^{\mathrm{rd}}$QT) framework~\cite{prosen2008third, prosen2011open}, which 
provides a powerful and systematic method for solving quadratic open quantum systems.
After mapping the spin chain to fermions via the Jordan--Wigner transformation, the Liouvillian becomes 
bilinear in Majorana operators and can be written as a quadratic form in Liouville space~\cite
{prosen2012diffusive}. In this representation, the dynamics of the Majorana operators is governed by a 
non-Hermitian single-particle matrix, whose spectrum fully determines the relaxation properties~\cite
{denardis2018hydrodynamic}. The SS corresponds to the unique zero mode of the Liouvillian and is fully 
characterised by the covariance matrix
\begin{eqnarray}
C_{ab} &=& \langle w_a w_b \rangle_{\mathrm{SS}}  - \delta_{ab} \\
    &=& \mathrm{Tr}\!\left( w_a w_b \rho_{\mathrm{SS}} \right) - \delta_{ab},\nonumber
\end{eqnarray}
which satisfies a Sylvester equation of the form
\begin{equation}
\mathbf{X}^{T} C + C \mathbf{X} = -4\mathbf{M}_{i},
\end{equation}
where $\mathbf{X}$ encodes the coherent Hamiltonian dynamics and dissipative couplings, while $\mathbf{M}$ arises from the dissipative part 
induced by the jump operators. Solving this linear matrix equation yields the exact SS covariance matrix without requiring 
time evolution. For the present model with uniform incoherent spin loss and pumping, the solution reproduces a completely 
factorized steady state, diagonal in the $\sigma^z$ basis, consistent with the analytical expression in Eq.~\eqref{eq:rho_NESS}. 
While static correlations vanish identically in the SS, the $3^{\mathrm{rd}}$QT framework provides the natural starting point 
for computing dynamical correlations by treating excitations on top of this Gaussian steady state.
The longitudinal spin operator,
\begin{equation}
\sigma_l^z = - i w_{2l-1} w_{2l},
\end{equation}
is bilinear in Majorana fermions, allowing time-dependent correlation functions to be expressed solely in terms of two-point correlators.
The longitudinal spin correlation function under full Lindbladian evolution can be computed using the third-quantization 
framework.

\begin{figure}[tbh!]
  \begin{center}
   \includegraphics[width=0.60\columnwidth]{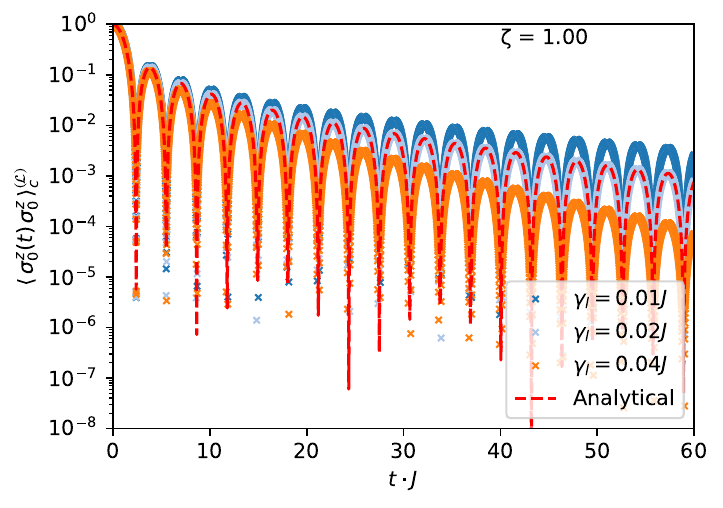}
   \caption{Longitudinal autocorrelation at the origin under Lindbladian dynamics for $\zeta=1$, $\Delta=0$, and several $\gamma_l$'s. Symbols: $3^{\mathrm{rd}}$QT; dashed line: exact analytical result Eq.~\eqref{eq:S0zz_Lindblad_XX} for 
   $\gamma_l = 0.02 J$. The signal undergoes an exponential decay $e^{-\Gamma t}$ modulated by $J_0^2(Jt)$ oscillations.
    See Appendix~\ref{sec:3rdQT} for details.}\label{fig:S0_zz_3rdQT}
  \end{center}
\end{figure}

The problem reduces to evaluating the time evolution of bilinear Majorana operators on top of the nonequilibrium 
steady state (see Appendix~\ref{sec:3rdQT}). Within $3^{\mathrm{rd}}$QT, the dynamics of Majorana correlators is fully captured by 
the time-dependent covariance matrix $C(t)$. The initial condition for $C(t)$ is fixed by inserting the operator $w_{2l'-1}w_{2l'}$
into the steady-state density matrix, corresponding to the preparation induced by the second spin operator in the correlator.
Owing to the Gaussian nature of $\rho_{\mathrm{SS}}$, the resulting four-point Majorana expectation values entering the initial condition are evaluated exactly using Wick's theorem. 
Once $C(t)$ is obtained, the longitudinal spin correlation function follows directly as
\begin{equation}
\average{\sigma_l^{z}(t)\,\sigma_{l'}^{z}}^{(\mathcal{L})}
=
-\,\mathrm{Tr}\!\left\{ w_{2l-1} w_{2l}\,\tilde\rho_{l'}(t)\right\}
=-\,C^{(l')}_{2l-1,2l}(t),
\end{equation}
where $\tilde\rho_{l'}(t)=e^{-i\mathcal{L}t}\big[w_{2l'-1}w_{2l'}\,\rho_{\mathrm{SS}}\big]$ is the (generally unnormalized) auxiliary density matrix associated with the operator insertion, and $C^{(l')}(t)$ denotes its covariance matrix.
This approach yields the full spatiotemporal structure of longitudinal correlations, revealing ballistic spreading inherited 
from the underlying XX dynamics, combined with an overall exponential decay set by the dissipative rates. The detailed 
derivation of the initial conditions and the explicit form of the covariance matrix evolution are presented in Appendix~\ref{sec:3rdQT}.

The SS covariance matrix is fully determined by the fugacity $\zeta$, with elements
\begin{equation}
C_{2l-1, 2l}^{\mathrm{SS}} = - C_{2l, 2l-1}^{\mathrm{SS}} = -i \frac{1-\zeta}{1+\zeta},
\end{equation}
and all other elements zero.

Figure~\ref{fig:S0_zz_3rdQT} shows the longitudinal autocorrelation at the origin
$\average{\sigma_0^{z}(t)\,\sigma_{0}^{z}}^{(\mathcal{L})}$
at $\zeta=1$, and a comparison between the exact analytical result Eq.~\eqref{eq:S0zz_Lindblad_XX} and the $3^{\mathrm{rd}}$QT calculation, demonstrating perfect agreement. The signal exhibits an overall 
exponential decay $\sim e^{-\Gamma t}$ modulated by coherent Bessel-function oscillations $J_0^2(Jt)$, 
reflecting the interplay of coherent ballistic propagation and incoherent damping. The same behavior is 
observed at other fugacities (not shown), with the exponential rate $\Gamma$ set purely by the total 
incoherent rate and independent of $\zeta$. 

\subsection{Interacting case, $\Delta\neq 0$: TEBD results}\label{sec:Lindblad_Delta_nonzero}

We now extend the analysis to the genuinely interacting regime $\Delta\neq 0$ under full
Lindbladian dynamics, where coherent XXZ exchange competes with local incoherent spin-flip
processes. The central question is whether the light-cone structures and their universality classes established in the unitary
problem survive the onset of dissipation, even at strong loss rates~\cite{alba2021spreading, znidaric2011spin}.

The main finding is that the two-regime picture established analytically for the free-fermion
limit — Eq.~\eqref{eq:Szz_Lindblad_XX} and the crossover at $t^{*}$ — extends 
to the full interacting case. Even at strong loss rates $\gamma_l\sim 0.1J$, both
the topology of the light cone and the universality class of its propagating front are fully
preserved~\cite{moca2023kardar}. The same two-regime structure governs the interacting dynamics,
but with the bare $\Gamma$ replaced by an effective damping rate $\Gamma_{\rm eff}^{(z)}(\Delta,\zeta)$
that acquires corrections from both the anisotropy $\Delta$ and the fugacity $\zeta$ through the
coherent-dissipative interplay and sets the actual crossover time scale $t\sim t^{*(z)}_{\rm eff}$.
For $t\ll t^{*(z)}_{\rm eff}$, the dynamics is dominated by coherent spin transport and the
Lieb-Robinson causal front builds up as in the purely unitary problem: ballistic, KPZ superdiffusive,
or diffusive depending on $\Delta$, with dissipation entering only as an overall multiplicative envelope.
For $t\gtrsim t^{*(z)}_{\rm eff}$, the exponential envelope becomes dominant and correlations are
uniformly suppressed regardless of spatial structure, so the light cone vanishes,
in the same way as in the free-fermion case. 
This robustness reflects a fundamental separation of
scales: the Lieb-Robinson velocity~\cite{lieb1972finite, bravyi2006lieb} sets the causal front and
is controlled by the coherent exchange $J$, while dissipation acts as a purely local decoherence
channel that shortens the lifetime of excitations.
\begin{figure*}[tbh!]
  \begin{center}
  \includegraphics[width=0.32\textwidth]{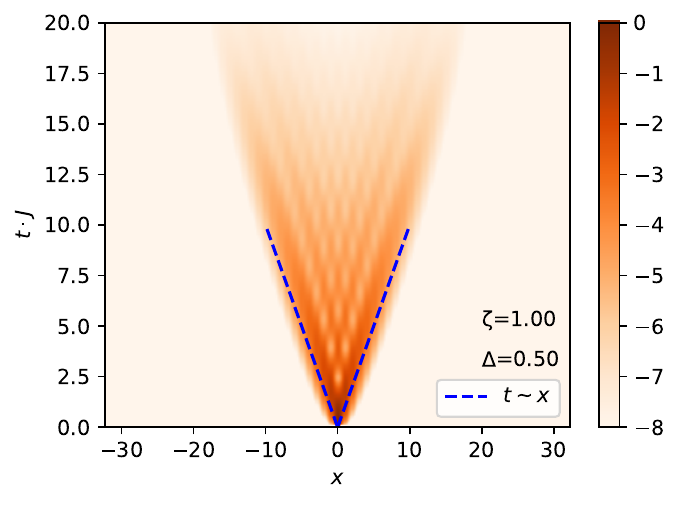}
  \includegraphics[width=0.32\textwidth]{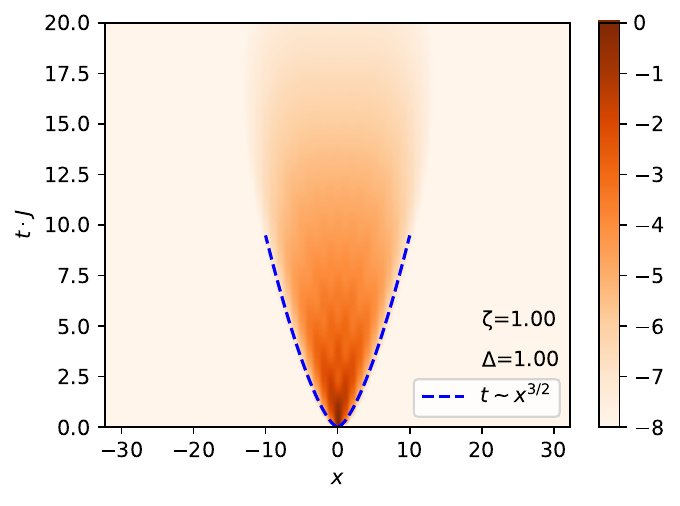}
  \includegraphics[width=0.32\textwidth]{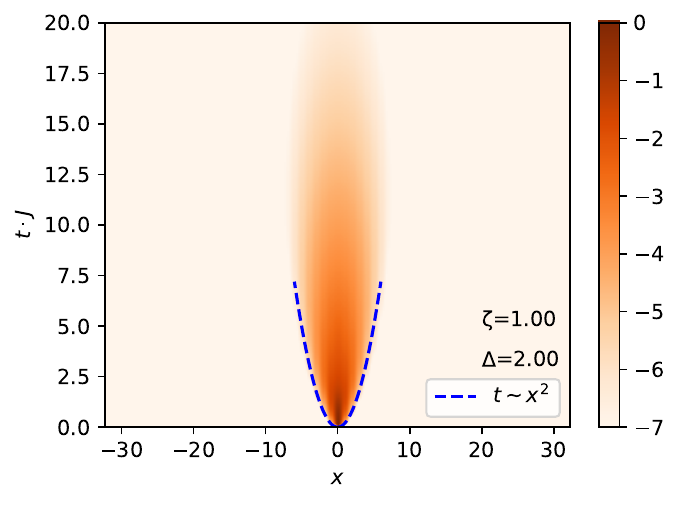}
  \includegraphics[width=0.32\textwidth]{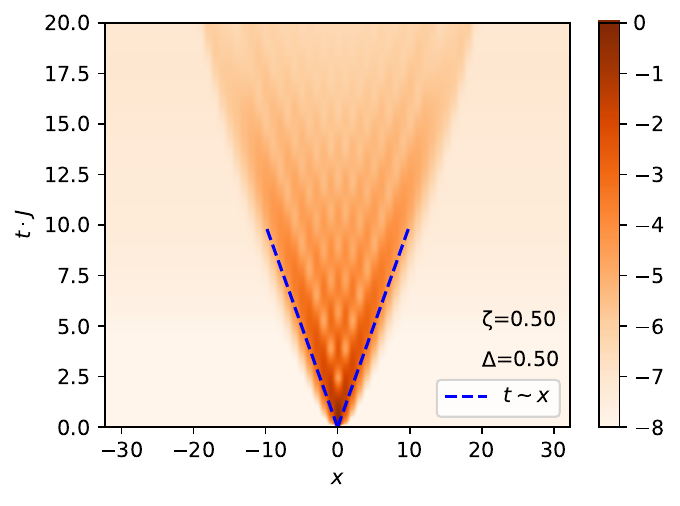}
  \includegraphics[width=0.32\textwidth]{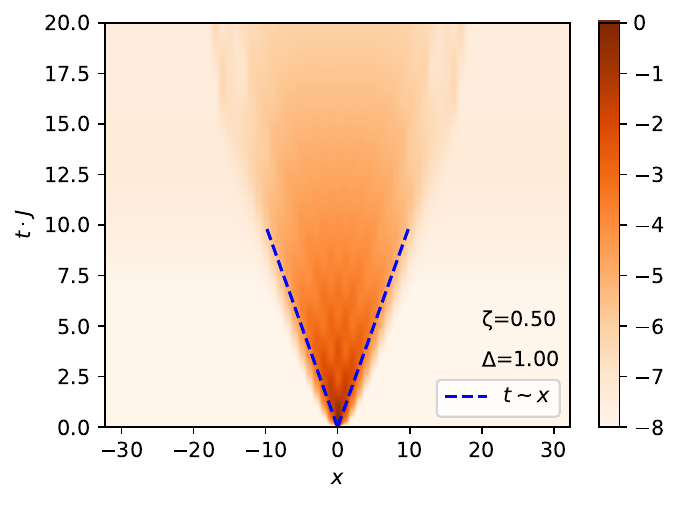}
  \includegraphics[width=0.32\textwidth]{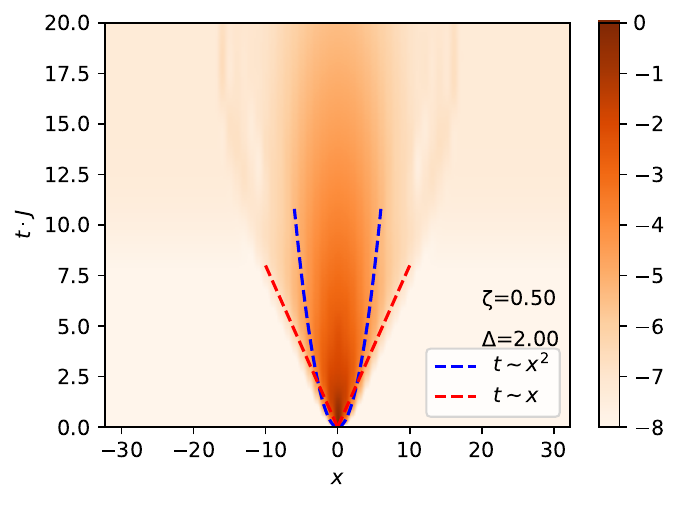}	
   \caption{Time evolution of the longitudinal autocorrelation $\langle \sigma_x^z(t)\,\sigma_0^z \rangle_{c}^{(\mathcal{L})}$ for
   several values of $\Delta$ and fugacity $\zeta$ (indicated in each panel). 
   The data is displayed on a logarithmic color scale.
   Dashed lines mark the light-cone boundaries at early times $t\lesssim t^{*(z)}_{\rm eff}$. The dynamical exponent $z$ governing the cone
   is ballistic ($z=1$) for $\Delta<1$, superdiffusive ($z=3/2$) at $\Delta=1$, and diffusive ($z=2$) for $\Delta>1$.
   For $\Delta=2.0$ and $\zeta=0.5$, two distinct fronts are visible, signaling coexistence of ballistic and diffusive modes. The loss rate is fixed to $\gamma_l=0.1J$.}
  \label{fig:S_zz_Delta_panels_Lindblad}
  \end{center}
\end{figure*}
\begin{figure}[tbh!]
  \begin{center}
   \includegraphics[width=0.49\columnwidth]{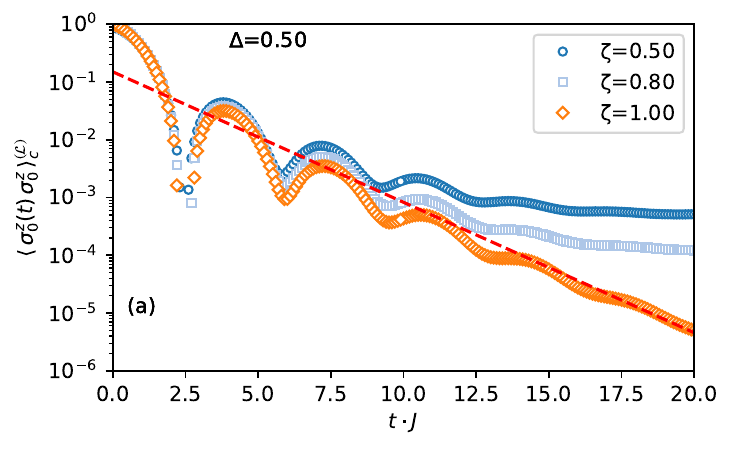}
   \includegraphics[width=0.49\columnwidth]{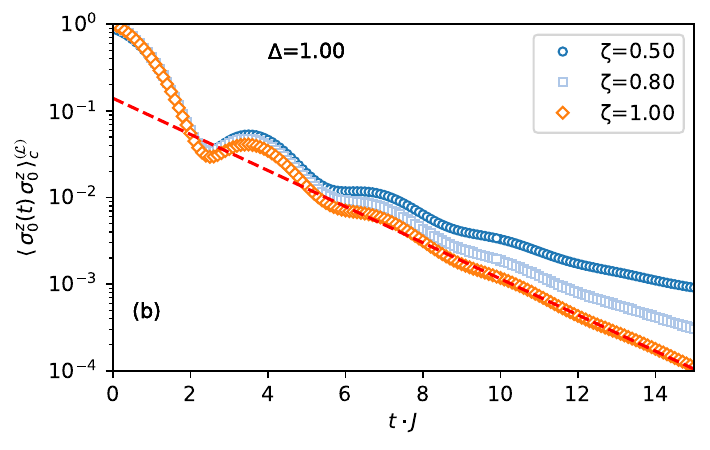}

   \caption{ Time evolution of the longitudinal correlation at 
   the central site in the presence of dissipation for $\Delta=0.5$ (a) and $\Delta=1$ (b)
   and different values of fugacity $\zeta$, displaying an exponential decay modulated by oscillations.
   The loss rate is fixed to $\gamma_l = 0.2J$. Dashed red lines indicate the exponential envelope decay.}\label{fig:S0_zz_Delta_05}
  \end{center}
\end{figure}
Figure~\ref{fig:S_zz_Delta_panels_Lindblad} shows the spatiotemporal structure at 
various anisotropies $\Delta$'s and fugacities $\zeta=\{0.5, 1.0\}$'s, 
with a fixed loss rate $\gamma_l=0.1J$. 
Despite this strong dissipation, the density plot (logarithmic color scale) unambiguously reveals
similar lightcones as in the unitary case (Fig.~\ref{fig:S_zz_Delta_panels}), with the same dynamical exponents $z$ governing the causal front:
ballistic ($z=1$) for $\Delta<1$, superdiffusive ($z=3/2$) at $\Delta=1$, and diffusive ($z=2$) for $\Delta>1$.
For $\Delta=1.0$, $\zeta=1$ the causal
front scales as $t\sim x^{3/2}$, identical to the KPZ universality class with dynamical
exponent $z=3/2$~\cite{ljubotina2019kardar, ilievski2018superdiffusion, krajnik2020kardar}.
However, the overall signal decays exponentially in time while the causal boundary retains its
non-trivial superdiffusive shape throughout the entire simulation window~\cite{moca2023kardar, bulchandani2021kardar}.

Figure~\ref{fig:S0_zz_Delta_05} examines the on-site longitudinal autocorrelation
$\average{\sigma_0^{z}(t)\,\sigma_{0}^{z}}^{(\mathcal{L})}$ for two representative anisotropies,
$\Delta=0.5$ (upper panel, easy-plane) and $\Delta=1$ (lower panel, isotropic), at several
fugacities $\zeta$. In both panels the signal exhibits an exponential decay modulated by coherent
oscillations — in striking contrast to the unitary case where the envelope follows an algebraic
power law~\cite{castroalvaredo2016emergent, bertini2016transport}. This behavior is the direct
interacting analogue of Eq.~\eqref{eq:Szz_Lindblad_XX}: for $t\ll t^{*(z)}_{\rm eff}$ transient
oscillations tied to the coherent transport mode are visible, while for $t\gtrsim t^{*(z)}_{\rm eff}$
the signal crosses over to a pure exponential decay. The effective rate $\Gamma_{\rm eff}^{(z)}(\Delta,\zeta)$
— visible as the slope on the semi-logarithmic scale — depends on both $\Delta$ and $\zeta$: 
it acquires an interaction-induced
corrections at finite $\Delta$, and varying $\zeta$ further shifts the crossover time~\cite{znidaric2019coexistence,deNardis2019anomalous}.

\section{Lindbladian evolution, Transverse correlations}\label{sec:transversal_Lindblad}

We now turn to the transverse autocorrelation at the central site under Lindbladian evolution,
$\average{\sigma_0^{x}(t)\,\sigma_{0}^{x}}^{(\mathcal{L})}$.
This quantity is considerably harder to access analytically than its longitudinal counterpart.
For the longitudinal channel at $\Delta=0$, the key simplification is that $\sigma_l^z = -iw_{2l-1}w_{2l}$ is
bilinear in Majorana operators, so the $3^{\mathrm{rd}}$QT framework provides a closed equation of motion for the covariance matrix.
The transverse operator $\sigma_l^x$, by contrast, involves a semi-infinite Jordan--Wigner string of occupation
number operators that makes it inherently nonlocal in the fermionic language.
Consequently, the equation of motion for the associated operator expectation value does not close at the two-point level, and
the $3^{\mathrm{rd}}$QT approach is not directly applicable.
Similarly, the Pfaffian method that evaluates $\average{\sigma_0^{x}(t)\,\sigma_{0}^{x}}^{(\mathcal{H})}$ exactly 
in the free-fermion case relies on the Wick factorizability of the Gaussian unitarily-evolved state;
under Lindbladian evolution the steady state remains Gaussian, but the propagation of the Jordan--Wigner string
no longer preserves Gaussianity~\cite{eisler2011crossover, eisler2014entanglement}, so that the Pfaffian representation
breaks down.
We therefore rely entirely on vectorised TEBD for all transverse correlators under dissipative dynamics.

Physically, the incoherent spin-flip processes introduce a spatially homogeneous decoherence channel that
suppresses off-diagonal elements of the density matrix, resulting in an overall exponential decay of the transverse
autocorrelation at an effective rate $\sim\Gamma_{\rm eff}^{(x)}(\Delta,\zeta)$.
At the same time, the coherent part of the Hamiltonian continues to drive oscillatory dynamics set by the
dispersion bandwidth $J$.
The interplay of these two processes results in a characteristic pattern: an exponentially decaying envelope
modulated by coherent oscillations whose frequency is set by $J$ and whose damping is governed by $\Gamma_{\rm eff}^{(x)}(\Delta,\zeta)$.

\begin{figure}[tbh!]
  \begin{center}
   \includegraphics[width=0.6\columnwidth]{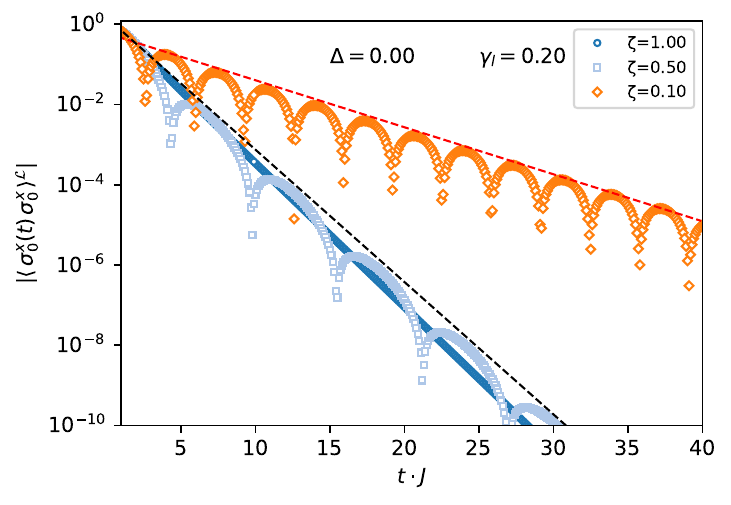}
   \caption{Transverse autocorrelation $\average{\sigma_0^{x}(t)\,\sigma_0^{x}}^{(\mathcal{L})}$ at $\Delta=0$,
   $\gamma_l=0.20J$, and several fugacities $\zeta$ (TEBD).
   The envelope decays exponentially at effective rate $\Gamma_{\rm eff}^{(x)}(0,\zeta)$ while the oscillation pattern and overall
   amplitude are controlled by the steady-state filling $\bar n = \zeta/(1+\zeta)$.}
   \label{fig:S0_xx_z}
  \end{center}
\end{figure}

\begin{figure}[tbh!]
  \begin{center}
   \includegraphics[width=0.60\columnwidth]{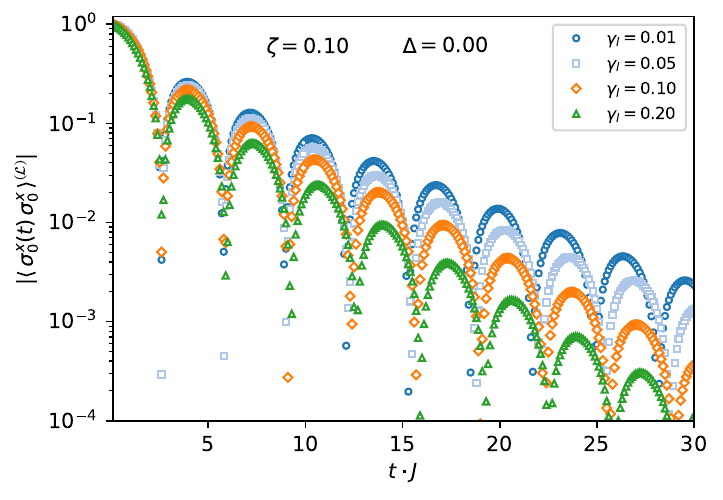}
   \caption{Transverse autocorrelation $\average{\sigma_0^{x}(t)\,\sigma_0^{x}}^{(\mathcal{L})}$ at $\Delta=0$,
   $\zeta=0.10$, and several loss rates $\gamma_l$ (TEBD).
   The rate $\gamma_l$ increases the exponential decay while leaving the coherent oscillation
   frequency set by $J$ unchanged, confirming that $\Gamma_{\rm eff}^{(x)}$ and the
   coherent exchange energy $J$ enter the dynamics independently.}
   \label{fig:S0_xx_gamma}
  \end{center}
\end{figure}

\begin{figure}[tbh!]
  \begin{center}
   \includegraphics[width=0.60\columnwidth]{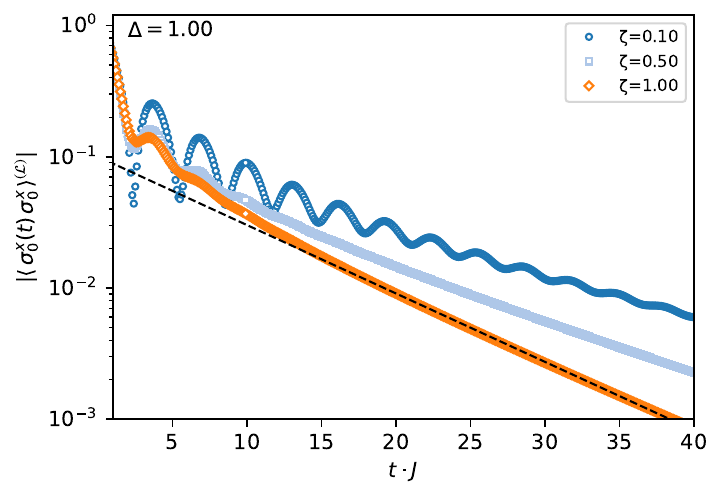}
   \caption{Transverse autocorrelation $\average{\sigma_0^{x}(t)\,\sigma_0^{x}}^{(\mathcal{L})}$ at the isotropic
   point $\Delta=1$ for several loss rates $\gamma_l$ (TEBD).
   Dissipation imposes an exponential suppression $\sim e^{-\Gamma_{\rm eff}^{(x)}(\Delta,\zeta)\,t}$, with the effective
   rate $\Gamma_{\rm eff}^{(x)}$ depending on both $\gamma_l$ and $\zeta$.}
   \label{fig:S0_xx_tebd_Delta1}
  \end{center}
\end{figure}

Figure~\ref{fig:S0_xx_z} illustrates the fugacity dependence of the transverse autocorrelation at $\Delta=0$ and
fixed loss rate $\gamma_l=0.2J$.
For all values of $\zeta$ the signal decays exponentially with effective rate $\Gamma_{\rm eff}^{(x)}(0,\zeta)$, reflecting the fact that the decoherence rate of the transverse channel is influenced by the background filling even at $\Delta=0$.
The oscillation pattern is strongly $\zeta$-dependent, with the most pronounced oscillations appearing at small $\zeta$. This is because the transverse channel is most coherent at low filling, where the system is close to the fully-polarized state and the Jordan--Wigner string has minimal effect on the dynamics.
As $\zeta$ increases, the filling approaches half-filling and the transverse channel becomes more strongly affected by the nonlocal string, leading to a suppression of coherent oscillations and a more rapid decay of the signal. For $\zeta=1$ (half-filling), the oscillations are almost completely washed out, and the signal decays rapidly without a clear oscillatory pattern.

Figure~\ref{fig:S0_xx_gamma} addresses the role of the dissipation rate at fixed $\zeta=0.10$ and $\Delta=0$.
As expected, increasing $\gamma_l$ leads to a progressively faster exponential decay, with $\Gamma_{\rm eff}^{(x)}$ growing linearly with increasing $\gamma_l$.
At the same time, the oscillation frequency — set by the exchange energy $J$ — remains unchanged across all
curves, demonstrating that $\Gamma_{\rm eff}^{(x)}$ and $J$ enter the transverse correlator through separate
channels: dissipation modulates the amplitude while the coherent exchange governs the phase.

Figure~\ref{fig:S0_xx_tebd_Delta1} extends the analysis to the isotropic point $\Delta=1$, where the competition
between the KPZ superdiffusion and dissipation is most interesting.
In the unitary case, the $SU(2)$ symmetry of the Heisenberg chain ties the transverse and longitudinal channels
together and enforces a $t^{-2/3}$ power-law decay of the transverse autocorrelation, as established in 
Fig.~\ref{fig:S_xx_Delta_1}.
For $\Delta=1$ under Lindbladian dynamics, two distinct time regimes are apparent.
At early times $t\ll t^{*(x)}_{\rm eff}$, the signal is consistent with the KPZ envelope, reflecting the fact that for
$t\Gamma_{\rm eff}^{(x)}\ll 1$ the coherent dynamics dominates and the effect of dissipation is merely a small overall prefactor
$e^{-\Gamma_{\rm eff}^{(x)} t}\approx 1$.
At later times $t\gtrsim t^{*(x)}_{\rm eff}$, the exponential suppression from the bath takes over and the signal
crosses over to a purely exponential decay, with the crossover time $t^{*(x)}_{\rm eff}$ shifting to earlier values as $\gamma_l$ is
increased.
This behavior demonstrates that bulk dissipation, while leaving the short-time KPZ dynamics intact, eventually
overwhelms the algebraic relaxation by imposing its own faster time scale $\sim t^{*(x)}_{\rm eff}(\Delta,\zeta)$.

\section{Conclusions}\label{sec:conclusions}

We have studied dynamical spin correlations in an open XXZ chain subject to uniform local spin-loss and pumping.
Although the nonequilibrium steady state is a featureless, correlation-free product state set entirely by the gain-loss balance, rich dynamics emerges once a local spin disturbance is imprinted on it.

Under purely coherent evolution the longitudinal channel reproduces the known hierarchy of spin-transport universality classes: ballistic spreading in the 
easy-plane phase, Kardar--Parisi--Zhang (KPZ) superdiffusion at the isotropic Heisenberg point, and diffusive relaxation in the easy-axis phase.
A finite longitudinal magnetization of the initial state superimposes an additional ballistic channel on top of these universality classes due to finite magnon density.
The transverse channel is qualitatively distinct: it decays rapidly without a light cone for easy-plane anisotropy, 
inherits KPZ scaling at the isotropic point, and supports a sharp ballistic wavefront for easy-axis anisotropy. Moreover, it is also largely initial longitudinal 
magnetization independent.
Notably, KPZ universality at the isotropic point persists for all simulated fugacities, indicating that it is protected by the underlying $SU(2)$ symmetry 
rather than by a particular filling.

Under Lindbladian evolution, we found an effective crossover time scale: $t^{*}_{\rm eff} \sim 1/\Gamma_{\rm eff}$ that controls the dynamics. 
The rate $\Gamma_{\rm eff}$ reduces to the bare decoherence rate $\Gamma=\gamma_l+\gamma_p$ in the noninteracting limit $\Delta=0$, but acquires distinct $\Delta$- and $\zeta$-dependent corrections in the presence of interactions, reflecting the different sensitivity of each channel to the background filling and anisotropy.
For $t\ll t^{*}_{\rm eff}$ the dynamics is effectively unitary: the shape of the light cone, the associated dynamical exponents, and the universality class are all fully preserved.
For $t\gtrsim t^{*}_{\rm eff}$ dissipation dominates and correlations decay 
exponentially, so the lightcone structure is losing its practical meaning. This reflects a clean 
separation of scales in which coherent exchange governs causal propagation while dissipation acts as a 
local decoherence channel that shortens the lifetime of excitations. 
These results suggest that measurements of dynamical spin correlations in quantum simulation platforms 
can reliably identify the transport universality class even in the presence of environmental decoherence, provided the observation time is shorter than the effective crossover scale.

\begin{acknowledgments}
This work received financial support from CNCS/CCCDI-UEFISCDI, under projects number PN-IV-P1-PCE-2023-0159 and PN-IV-P1-PCE-2023-0987, and by the ``Nucleu'' Program within the PNCDI 2022-2027, Romania, 
carried out with the support of MEC, project No. 27N/03.01.2023, component project code PN 23 24 01 04  and also by ``Nucleu'' Grant No. 30N/2023.
We acknowledge the Digital Government Development and Project
Management Ltd.~for awarding us access to the Komondor HPC facility based in Hungary.
This work was also supported by the National Research, Development and Innovation Office - NKFIH  Project No. K142179 and 
by the HUN-REN Hungarian Research Network through the Supported Research Groups
Programme, HUN-REN-BME-BCE Quantum Technology Research Group (TKCS-2024/34).
\end{acknowledgments}

\appendix

\section{Majorana correlations}\label{sec:Majorana}
Here we present the time-dependent Majorana correlations for the 
noninteracting XX chain, which form the building blocks for the 
Pfaffian evaluation of transverse spin correlators. Explicitly, we have
\begin{eqnarray}
\average{\eta_{2l-1}(t)\eta_{2l'-1}(t)} &=& \sum_{j} J_{j-l}(\Omega t)J_{j-l'}(\Omega t) \Big \{ i^{\,l-l'}\bar{n}  + i^{\,l'-l}(1-\bar{n}) \Big\},
\label{eq:Majorana_11}
\end{eqnarray}

\begin{eqnarray}
\average{\eta_{2l-1}(t)\eta_{2l'}(t)} &=& i \sum_{j} J_{j-l}(\Omega t)J_{j-l'}(\Omega t) \Big \{ -i^{\,l-l'}\bar{n}  + i^{\,l'-l}(1-\bar{n}) \Big\},
\end{eqnarray}

\begin{eqnarray}
\average{\eta_{2l}(t)\eta_{2l'-1}(t)} &=& -\average{\eta_{2l-1}(t)\eta_{2l'}(t)},\nonumber \\
\average{\eta_{2l}(t)\eta_{2l'}(t)} &=& \average{\eta_{2l-1}(t)\eta_{2l'-1}(t)}.
\end{eqnarray}

Similarly, for unequal-time correlators involving the initial Majorana operators, we have
\begin{eqnarray}
\average{\eta_{2l-1}(t)\eta_{2l'-1}} &=& J_{l'-l}(\Omega t)\,\Big \{ i^{\,l-l'}\bar{n}  + i^{\,l'-l}(1-\bar{n}) \Big\},\nonumber \\
\average{\eta_{2l-1}(t)\eta_{2l'}} &=& i J_{l'-l}(\Omega t)\,\Big \{ -i^{\,l-l'}\bar{n}  + i^{\,l'-l}(1-\bar{n}) \Big\}, \nonumber \\
\average{\eta_{2l}(t)\eta_{2l'-1}} &=& -\average{\eta_{2l-1}(t)\eta_{2l'}},\nonumber \\
\average{\eta_{2l}(t)\eta_{2l'}} &=& \average{\eta_{2l-1}(t)\eta_{2l'-1}}.
\label{eq:Majorana_12_unequal_time}
\end{eqnarray}

In these expressions the sums run over all lattice sites $j$ (thermodynamic limit), and the filling is
$\bar n=\langle c_l^\dagger c_l\rangle=\zeta/(1+\zeta)$.

In the thermodynamic limit the lattice sums appearing in the equal-time correlators simplify by the standard Bessel-function closure relation
\begin{equation}
\sum_{j=-\infty}^{\infty} J_{j-l}(x)\,J_{j-l'}(x)=\delta_{l,l'}.
\end{equation}
Applying this identity to Eq.~\eqref{eq:Majorana_11} (and the companion expressions above) yields the compact results
\begin{eqnarray}
\average{\eta_{2l-1}(t)\eta_{2l'-1}(t)} = \delta_{l,l'},\quad
\average{\eta_{2l-1}(t)\eta_{2l'}(t)} = i\,(1-2\bar n)\,\delta_{l,l'},\nonumber\\
\average{\eta_{2l}(t)\eta_{2l'-1}(t)} = -i\,(1-2\bar n)\,\delta_{l,l'},\quad
\average{\eta_{2l}(t)\eta_{2l'}(t)} = \delta_{l,l'}.
\end{eqnarray}

These expressions represent all possible combinations of two-operator expectation values between the time-evolved 
and initial Majorana 
operators. In the Pfaffian formulation of the transverse spin correlator, each entry of the covariance submatrix 
corresponds to one of these two-point averages. 

Notice that the time dependence of the correlation function enters only though  the unequal-time correlator in 
Eq.~\eqref{eq:Majorana_12_unequal_time} while   the equal-time correlators are time independent and reflect the local 
occupation of fermions in the steady state, which is set by the fugacity $\zeta$.

The transverse spin correlation function is then obtained as the Pfaffian of the matrix formed from 
these building blocks, automatically summing over all possible pairings with the correct fermionic signs.

\section{\texorpdfstring{$3^{\mathrm{rd}}$ QT calculation of the longitudinal spin correlation}{3rd QT calculation of the longitudinal spin correlation}}\label{sec:3rdQT}
The relation between the creation/annihilation operators and Majorana fermions is given by 
\begin{equation}
  c_l = \frac{1}{2}(w_{2l-1} - i w_{2l}), \quad c_l^{\dagger} = \frac{1}{2}(w_{2l-1} + i w_{2l}).
\end{equation}
which allows us to express the spin operator $\sigma_l^z$ in terms of Majorana fermions as
\begin{equation}
  \sigma_l^z = - i w_{2l-1} w_{2l}.
\end{equation}
We are interested in calculating the time-dependent longitudinal spin correlation function under Lindbladian evolution,
\begin{eqnarray}
\average{\sigma_l^{z}(t)\,\sigma_{l'}^{z}}^{(\mathcal{L})}
&=& 
\mathrm{Tr}\!\left\{
\sigma_l^{z}(t)\,\sigma	_{l'}^{z}\,\rho_{\mathrm{SS}}\right \} 
- \mathrm{Tr}\!\left\{
 (w_{2l-1} w_{2l})(t) \, (w_{2l'-1} w_{2l'})\,\rho_{\mathrm{SS}}
\right\} 
\end{eqnarray}	
To evaluate this quantity within third quantization, it is convenient to introduce the auxiliary (generally unnormalized)
operator
\begin{equation}
\tilde\rho_{l'}(t)=e^{-i\mathcal{L}t}\big[w_{2l'-1}w_{2l'}\,\rho_{\mathrm{SS}}\big],
\end{equation}
which is precisely the object appearing in the definition of the Lindblad correlator in Eq.~\eqref{eq:C_L_spin}.
In terms of $\tilde\rho_{l'}(t)$, the correlator becomes
\begin{equation}
\average{\sigma_l^{z}(t)\,\sigma_{l'}^{z}}^{(\mathcal{L})} = -\,\mathrm{Tr}\!\left\{ w_{2l-1} w_{2l}\,\tilde\rho_{l'}(t) \right\}.
\end{equation}
We therefore solve for the covariance matrix associated with $\tilde\rho_{l'}(t)$,
\begin{equation}
C_{ab}(t)=\mathrm{Tr}\!\left\{ w_a w_b\,\tilde\rho_{l'}(t)\right\}-\delta_{ab}\,\mathrm{Tr}\!\left\{\tilde\rho_{l'}(t)\right\},
\end{equation}
whose initial condition follows from the Gaussianity of $\rho_{\mathrm{SS}}$.
Explicitly,
\begin{equation}
C_{ab}(0)=\mathrm{Tr}\!\left\{ w_a w_b w_{2l'-1} w_{2l'} \,\rho_{\mathrm{SS}}\right\}-\delta_{ab}\,\mathrm{Tr}\!\left\{ w_{2l'-1} w_{2l'}\,\rho_{\mathrm{SS}}\right\}.
\label{eq:initial_C}
\end{equation}
Using Wick's theorem, we can evaluate the four-point correlator in the numerator as
\begin{eqnarray}
\mathrm{Tr}\!\left\{ w_a w_b w_{2l'-1} w_{2l'} \rho_{\mathrm{SS}}\right\} &=& 		
\langle w_a w_b \rangle_{\mathrm{SS}} \langle w_{2l'-1} w_{2l'} \rangle_{\mathrm{SS}}  - \langle w_a w_{2l'-1} \rangle_{\mathrm{SS}} \langle w_b w_{2l'} \rangle_{\mathrm{SS}} \nonumber \\
& & + \langle w_a w_{2l'} \rangle_{\mathrm{SS}} \langle w_b w_{2l'-1} \rangle_{\mathrm{SS}}.
\end{eqnarray}
With this initial condition, we can solve the equation for the time-dependent correlation matrix $C(t)$, 
\begin{equation}
\frac{d}{dt} C(t) = -2 \mathbf{X}^{T} C(t) - 2 C(t) \mathbf{X} - 8\,\mathbf{M}_{i}\,\mathrm{Tr}\!\left\{\tilde\rho_{l'}(t)\right\},
\end{equation}
where $\mathbf{X} = 2i \mathbf{H} +2\mathbf{M}_{r}$ and $\mathbf{M} $ is the matrix that incorporates the 
dissipative terms and are the same matrices appearing in the equation for the SS covariance matrix itself. 	
Finally evaluate the longitudinal spin correlation function as
\begin{equation}
\average{\sigma_l^{z}(t)\,\sigma_{l'}^{z}}^{(\mathcal{L})} = -\,C_{2l-1,2l}(t).
\end{equation}

\section{Pfaffian formulation}\label{sec:Pfaffian}
To illustrate the Pfaffian formulation explicitly, we consider the expectation value of a product of four Majorana operators,
$\langle \eta_{a}\eta_{b}\eta_{c}\eta_{d} \rangle$,
evaluated in a Gaussian state. According to Wick’s theorem, this correlator decomposes into a sum over all possible pairings,
\begin{equation}
\langle \eta_{a}\eta_{b}\eta_{c}\eta_{d} \rangle
=
\langle \eta_{a}\eta_{b} \rangle \langle \eta_{c}\eta_{d} \rangle
-
\langle \eta_{a}\eta_{c} \rangle \langle \eta_{b}\eta_{d} \rangle
+
\langle \eta_{a}\eta_{d} \rangle \langle \eta_{b}\eta_{c} \rangle .
\label{eq:wick4}
\end{equation}

The same result is obtained from the Pfaffian of the $4\times4$ covariance submatrix
\begin{equation}
\Gamma^{(a,b,c,d)} =
\begin{pmatrix}
0 & \Gamma_{ab} & \Gamma_{ac} & \Gamma_{ad} \\
-\Gamma_{ab} & 0 & \Gamma_{bc} & \Gamma_{bd} \\
-\Gamma_{ac} & -\Gamma_{bc} & 0 & \Gamma_{cd} \\
-\Gamma_{ad} & -\Gamma_{bd} & -\Gamma_{cd} & 0
\end{pmatrix},
\end{equation}
where $\Gamma_{ij} = \langle \eta_i \eta_j \rangle - \delta_{ij}$. The Pfaffian of an antisymmetric $4\times4$ matrix is given explicitly by
\begin{equation}
\operatorname{Pf}\!\left(\Gamma^{(a,b,c,d)}\right)
=
\Gamma_{ab}\Gamma_{cd}
-
\Gamma_{ac}\Gamma_{bd}
+
\Gamma_{ad}\Gamma_{bc}.
\end{equation}
Substituting the definition of $\Gamma_{ij}$ immediately reproduces Eq.~\eqref{eq:wick4}, 
demonstrating the equivalence between Wick’s theorem and the Pfaffian formulation for four operators.

\section{Vectorized TEBD calculation of correlation functions}\label{sec:TEBD}
The calculation of spin-spin correlation functions requires the time evolution of density matrices under either unitary or Lindbladian dynamics.
The vectorized TEBD method provides an efficient numerical framework to perform this task by leveraging the matrix product state (MPS) representation of mixed states, and the time evolution implemented using the Suzuki-Trotter decomposition of the evolution (super)operator.

Under the Choi-Jamio\l{}kowski  isomorphism~\cite{Choi1972,Jamiolkowski1972}, the density matrice $\rho$ are mapped to state vectors $\dket{\rho}$ as 
\begin{equation}
	\rho = \sum_{m,n} \rho_{mn} \ket{m}\bra{n} \mapsto \dket{\rho} = \sum_{m,n} \rho_{mn} \ket{m}\otimes\ket{n},
\end{equation} 
where $\ket{m}$ and $\ket{n}$ are basis states of the original Hilbert space.
The time evolution is governed by the Liouvillian superoperator $\mathcal{L}$, which acts on $\dket{\rho}$ as
\begin{equation}\label{eq:tebd}
	\frac{d}{dt} \dket{\rho} = -i \mathcal{L} \dket{\rho},
\end{equation}
where Eq.~\eqref{eq:tebd} is the vectorized form of the Lindblad master equation~\eqref{eq:lindblad}.
The superoperator $\mathcal{L}$ reads explicitly as
\begin{align}
	\mathcal L &= H\otimes \id - \id \otimes H^T 
	+ 
	\sum_{j=-L/2}^{L/2}
	i\frac{\gamma_l}{2}(2\sigma^-_j \otimes \sigma^-_j - \sigma_j^+\sigma^-_j \otimes \id - \id \otimes \sigma_j^+\sigma^-_j)\notag\\
	&\quad+\sum_{j=-L/2}^{L/2} i\frac{\gamma_p}{2}(2\sigma^+_j \otimes \sigma^+_j - \sigma_j^-\sigma^+_j \otimes \id - \id \otimes \sigma_j^-\sigma^+_j),
\end{align}
with $H$, the XXZ Hamiltonian~\eqref{eq:XXZ}.
For unitary evolution, the loss and pump rates $\gamma_l$ and $\gamma_p$ are set to zero.

The correlation functions are computed as expectation values in the doubled Hilbert space.
The spin-spin correlation functions are expressed as
\begin{equation}
	\average{\sigma_l^{\alpha}(t)\,\sigma_{l'}^{\beta}}
	= \mathrm{Tr}\!\left\{\sigma_l^{\alpha} e^{-iHt}
	\sigma_{l'}^{\beta}\rho_{SS} e^{iHt} \!\right\}
	=\dbra{\id}|(\sigma_l^{\alpha}\otimes\id)e^{-i\mathcal{L}t}
	(\sigma_{l'}^{\beta}\otimes\id)
	\dket{\rho_{SS}},
\end{equation}
with $\dket{\id}$ representing the vectorized identity operator in the doubled Hilbert space.
In simulations $l'$ is usually set to the central site, $l'=0$, while $l$ is varied across all the lattice sites.

The initial state $\dket{\rho_0}=(\sigma_{0}^{\beta}\otimes\id)\dket{\rho_{SS}}$ and the set of states
$(\sigma_l^{\alpha}\otimes\id)^\dag\dket{\id}$, for all sites $l$ in the lattice, are represented in matrix product state (MPS) form.
The dynamics of the $\dket{\rho_0}$ with the evolution operator $e^{-i\mathcal{L}t}$ is implemented using the TEBD algorithm, by performing a second-order Suzuki-Trotter decomposition with gates applied sequentially according to a brick-wall pattern. 
The time step is chosen to be sufficiently small $\delta t=0.01J^{-1}$ and bond dimension sufficiently large $\chi=256$ to ensure convergence of the results in the simulation time window. 
At a sequence of measurement time steps, the correlation function is evaluated by contracting the evolved MPS with the precomputed MPSs $(\sigma_l^{\alpha}\otimes\id)^\dag\dket{\id}$ to get the full spatiotemporal structure of the correlation function.

\bibliography{references}

\nolinenumbers

\end{document}